\documentclass[a4paper,11pt]{article}
\usepackage{jheppub}

\usepackage{mathtools}
\usepackage{orcidlink}
\usepackage[T1]{fontenc}
\usepackage{lmodern}
\usepackage[makeroom]{cancel}
\usepackage[titletoc]{appendix}
\usepackage{hyperref}
\usepackage{feynmf}
\usepackage{feyn}
\usepackage{csquotes}
\usepackage{slashed}
\usepackage{adjustbox}
\usepackage{xcolor}
\usepackage{dcolumn}
\usepackage{amsfonts}
\usepackage{indentfirst}
\usepackage{afterpage}
\usepackage{graphicx}
\usepackage{rotating}
\usepackage{amssymb}
\usepackage{braket}
\usepackage{mathrsfs}
\usepackage{subcaption}
\usepackage{appendix}
\usepackage{natbib}
\usepackage{float}
\usepackage{color}
\usepackage{soul}
\usepackage{microtype}
\usepackage{enumerate}
\usepackage{physics}
\usepackage{comment}
\usepackage{academicons}

\usepackage{tikz}
\usepackage{pgfplots}
\usetikzlibrary{shadings,intersections,decorations.pathmorphing}
\usetikzlibrary{automata,positioning}
\usepgfplotslibrary{colorbrewer}
\pgfplotsset{compat=newest}
\usepackage{fontawesome} 

\newcommand{\beq}{\begin{equation}}
\newcommand{\eeq}{\end{equation}}
\newcommand{\bea}{\begin{eqnarray}}
\newcommand{\eea}{\end{eqnarray}}
\newcommand{\be}{\begin{equation}} 
\newcommand{\ee}{\end{equation}}
\newcommand{\ba}{\begin{eqnarray}}
\newcommand{\ea}{\end{eqnarray}}
\newcommand{\alp}{\alpha}
\newcommand{\bet}{\beta}
\newcommand{\del}{\delta}

\newcommand{\kap}{\kappa}

\DeclareMathOperator{\arcsinh}{arcsinh}

\usepackage{graphicx}
\graphicspath{{./Figures/}}
\usepackage[margin=15pt,font={sf,it},labelfont={bf,sf},format=plain,justification=centerlast,width=.9\textwidth,skip=8pt]{caption}
\usetikzlibrary{matrix,calc,decorations.markings}
\usepackage{amsmath,amsthm}\usepackage[scaled=1.,ncf,vvarbb]{newtxmath}
\definecolor{DarkViolet}{RGB}{148,0,211}
\definecolor{azure(colorwheel)}{rgb}{0.0, 0.5, 1.0}
\definecolor{DarkBlue}{RGB}{0,0,154}
\definecolor{nicegreen}{RGB}{0,153,0}
\definecolor{nicered}{RGB}{255,66,66}

\newcommand{\mycolor}{DarkViolet}
\definecolor{tclr}{RGB}{148,0,211}

\hypersetup{colorlinks=true,bookmarksopen,bookmarksnumbered,citecolor={\mycolor},linkcolor={DarkBlue},urlcolor={DarkBlue},pdfstartview=FitH,anchorcolor={DarkBlue}}

\setstcolor{red}

\title{On the uplift of 4D wormholes in Braneworld models and their 5D structure}

\author[a]{Thomas D.~Pappas~\orcidlink{0000-0003-2186-357X}}
\emailAdd{thomas.pappas@physics.slu.cz}
\affiliation[a]{\href{https://www.slu.cz/phys/en/}{Research Centre for Theoretical Physics and Astrophysics,\\ Institute of Physics, Silesian University in Opava},\\ Bezručovo náměstí 13, CZ-746 01 Opava, Czech Republic}

\author[b]{and Theodoros Nakas\,\orcidlink{0000-0002-3522-5803}}
\emailAdd{theodoros.nakas@gmail.com}
\affiliation[b]{\href{https://ibs.re.kr}{Cosmology, Gravity, and Astroparticle Physics Group,\\ Center for Theoretical Physics of the Universe, Institute for Basic Science (IBS)},\\ Daejeon 34126, Korea}

\abstract{
Recent developments for the consistent embedding of general 4D static and spherically-symmetric spacetimes in arbitrary single-brane braneworld models in the form of the General Embedding Algorithm (GEA)~[\href{https://journals.aps.org/prd/abstract/10.1103/PhysRevD.109.L041501}{Phys.Rev.D 109 (2024) 4, L041501}], initiated the program of studying the bulk structure of braneworld wormholes. In this article, adopting a completely generic approach, we derive the general conditions that the metric functions of any braneworld spacetime must satisfy to describe a wormhole structure in the bulk. Particular emphasis is placed on clarifying the proper uplift of 4D spacetimes, expressed in terms of arbitrary radial coordinates on the brane, and we demonstrate the important role of the circumferential radius metric function $r(u)$ for the embedding.~To ensure applicability even when $r(u)$ is non-invertible, we develop an extended formulation of the GEA.
Additionally, the flare-out conditions for braneworld wormholes are presented for the first time and are found to differ from the case of flat extra dimensions. To illustrate the method, we first perform the uplift into both thin (Randall-Sundrum II) and thick braneworld models for four well-known 4D wormhole spacetimes: the effective braneworld wormhole solutions of Casadio-Fabbri-Mazzacurati and Bronnikov-Kim, the Simpson-Visser spacetime, and the Ellis-Bronnikov or~\enquote{anti-Fisher} solution. Subsequently, we study their bulk features by means of curvature invariants, flare-out conditions, energy conditions and embedding diagrams. Our analysis reveals that the assumption of a warped extra dimension has non-trivial implications for the structure of 5D wormholes.
}

\keywords{Wormholes, Modified Gravity, Braneworld models, Localized braneworld solutions.}

\begin{document}
\maketitle
\hypersetup{urlcolor={\mycolor}}
\flushbottom

\section{Introduction}
\label{Sec:Introduction}

Wormholes (WHs) are hypothetical, topologically non-trivial, spacetime structures with a \enquote{tunnel-like} shape that connect two asymptotic regions of the same or different universes and do not require the existence of extra dimensions for their formulation~\cite{Visser:1995cc}. The concept of a wormhole can be traced back to as early as 1916 with the work of~Flamm~\cite{Flamm:1916,2015GReGr..47...72F} and later on it re-emerged in the works of Einstein-Rosen~\cite{Einstein:1935tc} and Wheeler~\cite{Wheeler:1955zz} in the 1930s and 1950s respectively. However, these early WHs were non-traversable. The first known wormhole solution in General Relativity (GR) is the Ellis-Bronnikov~\cite{Bronnikov:1973fh,Ellis:1973yv} spacetime from the 1970s, that is sourced by a phantom (negative kinetic energy) scalar field. Nevertheless, it was not until the seminal works of Morris-Thorne~\cite{Morris:1988cz} and Visser~\cite{Visser:1989kh} on traversable WHs in the late 1980s that a significant resurgence of interest in the study of wormhole spacetimes emerged.

As the work of Morris and Thorne, and subsequent studies have revealed, WHs come with a number of problems such as the requirement of exotic forms of matter to prevent the WH from collapsing~\cite{Morris:1988cz,Morris:1988tu} and/or dynamical instabilities~\cite{Gonzalez:2008wd,Bronnikov:2011if}. Various attempts have been made in the literature to minimize the amount of the required exotic matter~\cite{Visser:1989kh} or even to avoid it completely in modified theories of gravity~\cite{Kanti:2011jz,Antoniou:2019awm}, however such solutions are often found to be unstable~\cite{Cuyubamba:2018jdl}. Recently, in the context of Einstein-Maxwell theory it was found that asymmetric WHs can indeed be obtained as solutions to the field equations without the need for exotic forms of matter~\cite{Blazquez-Salcedo:2020czn,Bolokhov:2021fil,Konoplya:2021hsm}, although a dynamical stability analysis for these solutions is still lacking. Another important theoretical challenge for the existence of WHs concerns their formation mechanism. In contrast to black holes (BHs) the formation of which can be modeled via gravitational collapse of matter~\cite{Misner:1973prb,Hawking:1973uf,Shapiro:1983du,Wald:1984rg}, WHs do not yet have a well-established formation mechanism. In fact, due to the topological difference between BHs and WHs, the formation of the later via gravitational collapse, seems to require new Physics, since, within the context of General Relativity, topological changes of spacelike sections entail causality violation~\cite{Geroch:1967fs}. To date, there is no known example of a fully satisfactory traversable Lorentzian WH that has been proven to not require exotic matter for its existence and being dynamically stable.

Despite the aforementioned challenges, wormhole structures cannot be ruled out and attempts to find observational evidence for their existence are ongoing~\cite{Bambi:2021qfo}. Such efforts are further motivated by the recent significant advancements of our observational facilities allowing us to probe the strong gravity regime in the gravitational~\cite{LIGOScientific:2016aoc} and electromagnetic spectrum~\cite{EventHorizonTelescope:2019dse}. Even though WHs are drastically different objects from BHs, their appearance need not always deviate significantly from that of a BH~\cite{Xavier:2024iwr} since there exist many examples in the literature of WHs acting as BH mimickers~\cite{Damour:2007ap,Bronnikov:2019sbx,Bronnikov:2021liv} and as such various methods to observationally distinguish between the two have been considered, see e.g.~\cite{DeFalco:2020afv,DeFalco:2021klh} and references therein

Even though much of the work on WHs has been performed in the familiar 4D spacetime, their study in the broader context of higher-dimensional settings has gained significant attention in recent years. Extra dimensions are motivated by a number of compelling reasons such as their crucial role in String Theory and in other approaches to quantum gravity, and their potential to resolving long-standing problems in particle Physics and Cosmology~\cite{Akama:1982jy,Rubakov:1983bz,Rubakov:1983bb,Arkani-Hamed:1998jmv,Arkani-Hamed:1998sfv,Antoniadis:1998ig,Randall:1999ee,Randall:1999vf}. In such theories, the extra dimensions are compactified or otherwise hidden from direct detection and thus our Universe appears to be effectively four-dimensional. In scenarios where the extra dimensions are flat, wormhole spacetimes provide simple, yet illuminating examples of how the existence of extra spacelike dimensions affects the basic properties of these exotic objects such as stability, energy conditions, and traversability~\cite{Torii:2013xba,Kuhfittig:2018voi}. Moreover, these flat extra-dimensional models act as stepping stones for studying more complicated frameworks.

In the~\emph{braneworld models}~\cite{Randall:1999ee, Randall:1999vf}, introduced in the late 1990s as a means to address the hierarchy problem of the huge discrepancy between the electroweak and gravity scale, the extra dimension is assumed to be warped. Such braneworld scenarios seem to have regained popularity in recent years since they were found to be useful for holography as well \cite{Maldacena:1997re, Gubser:1998bc, Witten:1998qj, Antonini:2019qkt, Geng:2021mic, Geng:2022dua}.
This seemingly simple assumption about the nature of the extra dimension however makes the study of higher dimensional objects a highly non-trivial task. Due to the difficulties in constructing consistent 5D objects in the presence of warped extra dimensions, studies of braneworld BHs and WHs have been performed mainly in terms of the effective brane field equations~\cite{Shiromizu:1999wj,Kanno:2002ia,Biswas:2023ofz}. Such an approach however, remains agnostic to the bulk content and provides no insight into the 5D structure of these objects~\cite{Kar:2015lma,Bronnikov:2019sbx,Biswas:2022wah}. In recent developments, a general embedding algorithm (GEA) that allows for the uplift on any static and spherically-symmetric brane metric to any single-brane branewrold model characterized by a non-singular and injective warp factor has been introduced in~\cite{Nakas:2023yhj} generalizing the embedding approach introduced in~\cite{NK1} for the Schwarzschild BH metric on the brane and the Randall–Sundrum II (RS-II) braneworld model~\cite{Randall:1999vf}. With the GEA it is now possible to perform a study of the complete 5D structure of braneworld WHs.

The article is organized as follows. In Sec.~\ref{Sec:2}, we define in the most general terms static and spherically symmetric WHs in spacetimes with an arbitrary number of \emph{flat} extra spacelike dimensions, and provide general conditions that the metric functions should satisfy in order for the geometry to have a wormhole topology. 
In Sec.~\ref{Sec:3}, we derive the general conditions that the metric functions of any five-dimensional single-brane braneworld model should satisfy to ensure that the bulk spacetime, with injective and non-singular warp function, exhibits a wormhole structure.
Under these conditions, we show that the bulk wormhole structure naturally induces a static and spherically symmetric wormhole on the brane.
We demonstrate that the uplift in the bulk can be performed in terms of an \emph{arbitrary} radial coordinate. 
In this regard, the embedding method presented in this article extends beyond the GEA formalism, and that is why it will be referred to as eGEA. 
In the last part of Sec. \ref{Sec:3}, we determine the general stress-energy tensor that is necessary to support solutions resulting from both GEA and eGEA, and discuss the energy conditions.
Subsequently, in Sec.~\ref{Sec:Ex}, we consider as examples various well-known 4D wormhole spacetimes, perform their uplift into the bulk for thin (RS-II) and thick braneworld models, and analyze their curvature invariants, flare-out and energy conditions, and their embedding diagram representations. 
Lastly, in Sec.~\ref{Sec:Conclusions}, we provide a summary of our findings and discuss future directions of research.

\textbf{Conventions and notation:} We adopt the $(-,+,+,+,+)$ signature for the metric and use geometrized units, namely $c=G=1$. Capital Latin letters will be employed to represent indices within the 5D bulk spacetime, while lowercase Greek letters will be used to denote indices related to the 4D brane. The total differentiation of a function $f$, with respect to its stated argument, will be denoted by $f'$. 
A subscript ${\rm d}$ in the line element $\mathrm{d}s_{\rm d}^2$ indicates that this metric constitutes a ${\rm d}$-dimensional hypersurface in the entire higher-dimensional spacetime, $\mathrm{d}s^2$.
The coordinate location of the throat in the entire higher-dimensional spacetime will be denoted by $R_\Join$.
The subscript ``$\rm{th}$'' on a radial coordinate also indicates the throat's location but refers to its projection onto lower-dimensional hypersurfaces.
We call Morris-Thorne (MT) metric frame the metric coordinate system in which the circumferential radius is identified with the radial coordinate. On the other hand, we call non-Morris-Thorne (nMT) the metric frame in which the previous condition does not hold. In the MT frame and for flat extra dimensions,~$R_\Join$ coincides with the physical radius of the throat. With a slight abuse of terminology we will extend the classification of a metric frame on the brane as MT or nMT also into the bulk, where the presence of the warp factor formally results in a nMT in all cases.
We use tilde over the $g_{\rm 00}=-\widetilde{f}$ and $g^{\rm 11}=\widetilde{h}$ metric functions in the nMT frame, to distinguish them from their MT versions which we denote as $f$ and $h$ respectively.

\section{Wormholes in D flat dimensions}
\label{Sec:2}

In terms of an arbitrary radial coordinate $u \in (-\infty,+\infty)$ the most general line element for a static spherically symmetric spacetime in ${\rm D=4+n}$ flat dimensions, with ${\rm n} \in \{0,\mathbb{N}\}$ being the number of extra dimensions, is given by~\footnote{Even though only two out of the three metric functions $\tilde{f}(u),\tilde{h}(u)$ and $r(u)$ in~\eqref{eq:ds2_nMT_unwarped} are independent due to the freedom in reparametrization of the radial coordinate, we here develop the formalism in the most general way possible.}
\beq
\dd s^2=-\widetilde{f}(u)\dd t^2+\frac{\dd u^2}{\widetilde{h}(u)}+r^2(u) \dd \Omega^2_{\rm D-2}\,,
\label{eq:ds2_nMT_unwarped}
\eeq
where the line element on the unit $({\rm D-2})$-sphere reads as
\beq
\dd \Omega^2_{{\rm D-2}}=\dd \vartheta_1^2+\sum_{i=2}^{{\rm D-2}} \prod_{j=1}^{i-1} \sin^2\vartheta_{j} \dd\vartheta_{i}^2 \,,
\eeq
with $\vartheta_i\in[0,\pi]$, $\forall\, i\in\{1,\ldots, {\rm D-3}\}$, and $\varphi\equiv \vartheta_{{\rm D-2}} \in[0,2\pi)$\,. The induced line element on a spacelike ($t=const.$) hypersurface of fixed radial coordinate $u=const=u_{\rm 0}$ in~\eqref{eq:ds2_nMT_unwarped} is given by 
\beq
\dd s_{\rm D-2}^2=r^2(u_{\rm 0})\dd\Omega^2_{\rm D-2}\,.
\eeq
The total surface area, or the $({\rm D-2})$-dimensional volume, to be more specific, enclosed by the line element~$\dd s_{\rm D-2}^2$ is determined by the following relation
\beq
S(u_{\rm 0})=\int_{0}^{2\pi}\int_0^\pi \cdots \int_0^\pi \left(\prod_{i=1}^{\rm D-3}\sqrt{g_{\vartheta_i\vartheta_i}}\, \dd\vartheta_i\right) \sqrt{g_{\varphi\varphi}}\, \dd\varphi \Bigg|_{u=u_{\rm 0}} =\frac{2\pi^{\rm (D-1)/2}}{\Gamma\left(\frac{\rm D-1}{2}\right)} r^{\rm D-2}(u_{\rm 0})\,.
\label{eq:total_area}
\eeq
The form of Eq.~\eqref{eq:total_area} reveals that it is the circumferential-radius metric function $r(u) \geqslant 0$, not $u$, that has the normal meaning of the radial variable. The radial coordinate $r$ is oftentimes also called~\emph{Schwarzschild} or \emph{curvature} radial coordinate since it corresponds to the radius of a sphere with radius $r$~\cite{Bronnikov:2012wsj}.

We wish to approach the definition of a wormhole metric described by~\eqref{eq:ds2_nMT_unwarped} in the most general way possible, and to this end we start with the minimal requirement of a~\enquote{tunnel-like} structure that connects two asymptotic regions of spacetime. The~\enquote{throat} of the WH, is defined as the narrowest part of that tunnel. Geometrically, such a structure in~\eqref{eq:ds2_nMT_unwarped} is characterized by the existence of a ${\rm (D-2)}$-dimensional spacelike hypersurface of constant radial coordinate with a minimal area at the throat~\cite{Hochberg:1997wp,Visser:1997yn}. Then, in terms of~\eqref{eq:total_area}, the coordinate location of the wormhole throat $u_{\rm th}$ is determined in the most general way, and for any ${\rm D}$, as the solution to the minimization conditions~\footnote{In general, it is possible to have more than one roots to~\eqref{eq:r0_cond_1}, in which case, the WH has multiple throats (if $r''(u_{\rm 0})>0$) or~\emph{equators} (if $r''(u_{\rm 0})<0$), see e.g.~\cite{Chew:2018vjp,Antoniou:2019awm,Chew:2021vxh}.}
\begin{alignat}{3}
&\left.\frac{\dd S(u)}{\dd u}\right\vert_{u=u_{\rm th}} &&=0 \quad \Rightarrow \quad \left.\frac{\dd r}{\dd u}\right|_{u=u_{\rm th}} &&=0\,,
\label{eq:r0_cond_1}\\[2mm]
&\left.\frac{\dd^2 S(u)}{\dd u^2}\right\vert_{u=u_{\rm th}} &&>0 \quad \Rightarrow \quad \left.\frac{\dd^2 r}{\dd u^2}\right|_{u=u_{\rm th}} &&>0
\label{eq:r0_cond_2}\,.
\end{alignat}
The physical radius of the throat, that is a gauge-invariant characteristic of the WH, is given by the circumferential radius metric function evaluated at $u_{\rm th}$, i.e.,
\beq
R_{\Join}\equiv r(u_{\rm th})\,.
\eeq
Notice that, even though the throat is located at a constant radial coordinate $u_{\rm 0}$, its physical radius $R_{\Join}$, need not be spherical when the geometry is axisymmetric, see e.g.~\cite{Teo:1998dp}.

We have thus far established a precise and fully general prescription to determine the coordinate location and radius of the wormhole throat. However, we need to impose further restrictions on the metric functions of~\eqref{eq:ds2_nMT_unwarped} for it to describe a traversable Lorentzian WH. The~\enquote{tunnel-like} topology connecting two asymptotic regions of spacetime located at $u \to \pm \infty$, in compliance with the minimality of the tunnel at the throat implies that the circumferential radius should be large compared to the minimum on both sides of the throat i.e.
\beq
r(u)\gg r(u_{\rm th})\,.
\label{eq:nMT_flare_out_flat_D}
\eeq
Finally, traversability requires the absence of event horizons everywhere in spacetime and this entails that the remaining two metric functions should satisfy
\beq
\widetilde{f}(u)>0\,,\quad \widetilde{h}(u)>0\,.
\label{eq:traversability_nMT}
\eeq

Another very commonly-used coordinate frame for the representation of wormhole spacetimes is the one  Morris and Thorne (MT) chose to work with in their seminal work~\cite{Morris:1988cz}. They defined wormhole geometries in a coordinate system with the circumferential radius as the radial coordinate, which, when generalized to ${\rm D}$ dimensions reads as
\beq
\dd s^2 =-f(r) \dd t^2 +\frac{\dd r^2}{h(r)}+r^2 \dd \Omega^2_{\rm D-2}\,.
\label{eq:ds2_MT_unwarped}
\eeq
The line element~\eqref{eq:ds2_MT_unwarped} is sometimes referred to in the wormhole literature as the~\enquote{MT frame}, and for the remainder of this article, we will adopt the same terminology; furthermore, we will call~\eqref{eq:ds2_nMT_unwarped} the~\enquote{non-MT frame} (nMT) in order to distinguish the two. Under the assumption of an invertible circumferential radius metric function $r(u)$ in~\eqref{eq:ds2_nMT_unwarped}, a straightforward correspondence between the MT and nMT frame representations of a WH can be established by means of a radial coordinate redefinition $u\to r=r(u)$. Furthermore, it is worth pointing out that in the case of asymmetric WHs w.r.t. the throat, the MT frame requires two sets of metric functions $\{f_{\pm}(r),h_{\pm}(r)\}$, with the subscripts $\pm$ introduced here to denote the metric functions on either side of the throat, which have to be matched smoothly across the throat in order for~\eqref{eq:ds2_MT_unwarped} to describe the complete spacetime. This is not necessary with the nMT frame which incorporates any throat asymmetry in the functional form of $r(u)$.

Contrary to WHs defined in the nMT frame~\eqref{eq:ds2_nMT_unwarped}, where the circumferential radius metric function $r(u)\geqslant0$ exhibits a minimum w.r.t. the radial coordinate $u$, in the MT frame~\eqref{eq:ds2_MT_unwarped} it does not, since it is identified with the radial coordinate $r\geqslant0$, and is in this sense a linear function of the radial coordinate. Consequently, it becomes obvious that different conditions to~\eqref{eq:r0_cond_1}-\eqref{eq:r0_cond_2} are necessary for the determination of the wormhole throat radius in this case. In the MT frame, a wormhole structure requires the existence of a lower bound in the range of $r$, i.e. $r \in [R_{\Join},+\infty)$, where $R_{\Join}>0$ is identified with the radius of the wormhole throat. Thus the appropriate condition for determining $R_{\Join}$ in the MT comes from the divergence of the $g_{rr}$ metric component which is equivalent to the surface area extremization condition (the analog of~\eqref{eq:r0_cond_1} in the nMT frame)
\beq
h(R_\Join)=0\,.
\label{eq:MT_u_th_general}
\eeq
Furthermore, to ensure that the metric has the topology of a WH, with the throat corresponding to the minimum, one has to also consider the so-called flare-out conditions evaluated at or near the location of the throat. For the metric~\eqref{eq:ds2_MT_unwarped}, and for any number of \emph{flat} dimensions these correspond to (analog of~\eqref{eq:r0_cond_2} in the nMT frame)
\beq
\frac{h'(r)}{2\left(h(r)-1\right)^2}>0\,.
\label{eq:MT_flare_out_flatD}
\eeq
The above relation can be easily obtained by following the same analysis presented in e.g.~\cite{Kim:2013tsa}.
Finally, traversability once again requires $f(r)>0$ everywhere (analog of \eqref{eq:traversability_nMT} in the nMT frame). 

One of the main objectives in this work, is the formulation of the corresponding general conditions on the metric functions of ${\rm 5D}$ WHs, in single-brane braneworld models, where the extra dimension is warped with an injective and non-singular warp factor.

\section{Extended General Embedding Algorithm for wormholes in 5D braneworld models}
\label{Sec:3} 

This section is divided into three parts. 
In the first part, we provide an in-depth overview of the General Embedding Algorithm (GEA) as introduced in~\cite{Nakas:2023yhj} and demonstrate its application to Morris-Thorne (MT) wormholes.
We show that transferring the domain of the 4D wormhole radial coordinate to the 5D radial coordinate is essential for consistently uplifting wormhole solutions to braneworld models.
Additionally, we derive the flare-out conditions for braneworld wormholes in general, highlighting the important role of the warp factor.

In the second part of this section, we perform the analysis in the non-Morris-Thorne (nMT) frame by introducing the extended formulation of the GEA, henceforth referred to as the extended General Embedded Algorithm (eGEA). In the special case where the circumferential radius metric function $r(u)$ of the seed brane metric is invertible, we demonstrate that the correspondence between the MT and nMT frames in 4D, is preserved by the uplifted wormhole. However, seed metrics with non-invertible $r(u)$ can be only uplifted into the bulk via the eGEA.

In the final part, we investigate the general characteristics of the stress-energy tensor necessary to support the braneworld geometries resulting from the GEA and eGEA. 
The analysis is done under the assumption that gravity and matter are decoupled from one another.
Lastly, we discuss the energy conditions.

\subsection{Morris-Thorne wormholes in the original formulation of the GEA}
\label{Sec:MT}

In this subsection, we provide an in-depth overview of the GEA as introduced in terms of the curvature radial coordinate $r$ in~\cite{Nakas:2023yhj}, extend the discussion provided therein, and highlight some of its important brane-localization properties. Subsequently, in line with the analysis of~Sec.~\ref{Sec:2}, we discuss in detail the general conditions that the metric functions of the GEA metric must satisfy in order for it to describe a wormhole structure in braneworld models.

The general line element for a flat 3-brane embedded in a warped $5$D spacetime with warp function $A(y)$ is given by
\beq
\dd s^2=e^{2A(y)}\eta_{\mu\nu}\dd x^\mu\dd x^\nu +\dd y^2=e^{2A(y)}\left(-\dd t^2+\dd r^2+r^2\dd \Omega^2_2 \right)+\dd y^2\,,
\label{eq:Warped_flat_original}
\eeq
where $\eta_{\mu\nu}$ is the four-dimensional Minkowski metric, and $y \in (-\infty,+\infty)$ is the coordinate parametrizing the extra dimension. The hypersurface $y=0$ represents our four-dimensional Universe, and is referred to as the~\emph{brane}. In the context of braneworld models, the warp function $A(y)$, satisfies $A(0)=0$ and its functional form is such that~\eqref{eq:Warped_flat_original} reduces to AdS$_5$ spacetime as $|y|\rightarrow+\infty$.
Additionally, in the overwhelming majority of braneworld models, the bulk is considered to be $\bold{Z}_2$-symmetric, namely $A(y)=A(-y)$.
The domains for the remaining coordinates are $t \in (-\infty,+\infty)$, $r\in[0,+\infty)$, $\vartheta \in [0,\pi]$ and $\varphi\in [0,2\pi)$. Regarding the spatial symmetries of the five-dimensional spacetime~\eqref{eq:Warped_flat_original}, it is axisymmetric w.r.t. $y$, while, the induced four-dimensional metric on any $y_0=const.$ slice is spherically symmetric w.r.t. $r$.

A method to generalize~\eqref{eq:Warped_flat_original} for a curved brane, proved to be a very challenging endeavor that required over two decades for its proper formulation \cite{NK1,NK2,Nakas:2023yhj}.~According to the intuitive approach followed by Chamblin, Hawking and Reall (CHR)~\cite{Chamblin:1999by}, the four-dimensional Minkowski metric in~\eqref{eq:Warped_flat_original} is replaced by a curved four-dimensional metric $g_{\mu\nu}$ leading to the 5D spacetime
\beq
\dd s^2=e^{2A(y)} g_{\mu\nu}\dd x^\mu\dd x^\nu +\dd y^2\,.
\label{eq:CHR_metric}
\eeq
Such an approach is however problematic since, despite inducing the four-dimensional metric $g_{\mu\nu}$ on the brane, the five-dimensional spacetime~\eqref{eq:CHR_metric}~is plagued with the undesirable general property of propagating features of the seed brane metric all the way to the AdS$_5$ boundary. For example, when $g_{\mu\nu}$ is that of a four-dimensional BH solution containing a curvature singularity at its center, the bulk geometry according to~\eqref{eq:CHR_metric}, corresponds to a black-string spacetime \cite{Tang:2022bcm, Rezvanjou:2017hox, Nakas:2020crd, Nakas:2019rod, Kanti:2018ozd}. Five-dimensional braneworld extensions of wormhole spacetimes via the CHR approach have also been considered in the literature, see e.g.~\cite{Sharma:2021kqb,Crispim:2024yjz}. More recently, it has been demonstrated that even in cases where $g_{\mu\nu}$ is free of curvature singularities, such as for the de Sitter metric~\cite{Nakas:2023yhj}, or even for wormhole spacetimes~\cite{Crispim:2024yjz}, the CHR bulk metric~\eqref{eq:CHR_metric} may yield curvature invariants that diverge at the boundary of AdS$_5$. 

\begin{figure*}[ht!]
\includegraphics[width=0.495\linewidth]{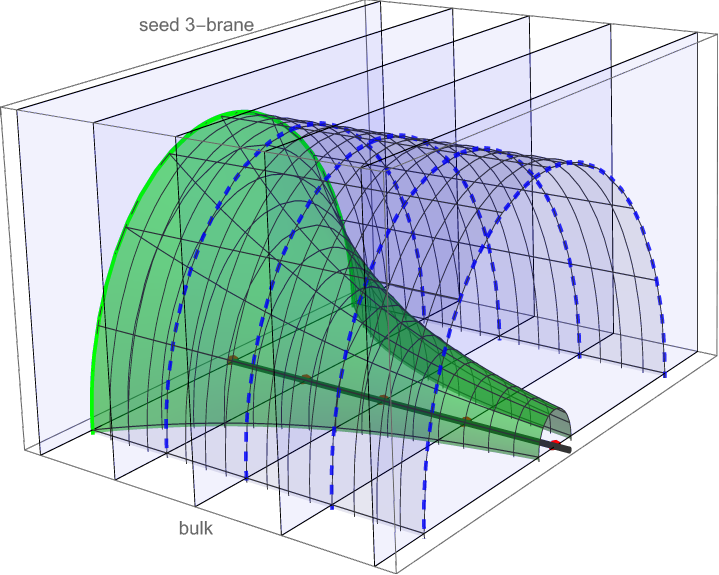}
\includegraphics[width=0.49\linewidth]{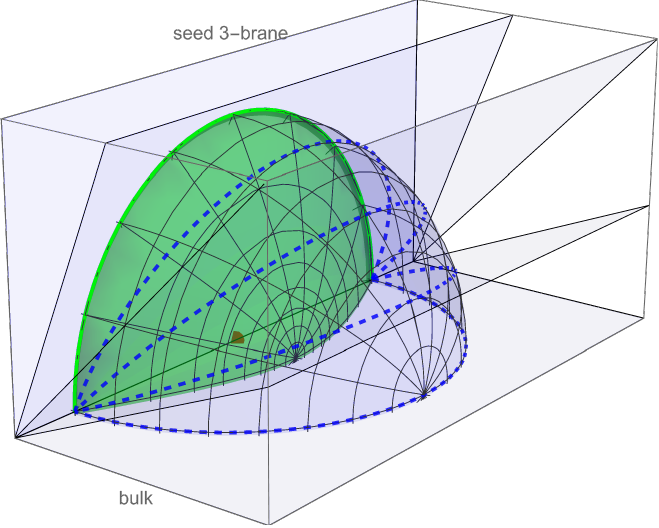}
\caption{The ${\rm 5D}$ bulk extension of the Schwarzschild BH in the RS-II model using the CHR method (left panel) and the GEA (right panel). The light-blue planes correspond to identical copies of the seed brane that continuously extend the geometry into the bulk prior to warping. The dashed blue curves correspond to the locations of the event horizon in each of the replica branes, while the red dots correspond to the locations of the curvature singularities. The green surfaces correspond to the ${\rm 5D}$ event horizon of the final warped bulk geometry. The CHR method yields a black string with the curvature singularity and the event horizon extending indefinitely into the bulk, while the GEA generates a BH with a finite-area event horizon and a curvature singularity that is localized strictly on the brane.}
\label{fig:CHR_vs_GEA_illustrated}
\end{figure*}

According however, to the recently proposed GEA~\cite{Nakas:2023yhj} that generalized the prescription introduced in~\cite{NK1}, all of the aforementioned drawbacks of the CHR method are resolved. The GEA-generated 5D geometries yield curvature invariants that are localized near the brane while reducing to AdS$_5$ asymptotically into the bulk (see~\cite{NK1,NK2,Neves:2021dqx,Nakas:2023yhj,Crispim:2024nou,Neves:2024zwi}). The reason for this, is the drastically different approach of the two methods in uplifting the brane metric into the bulk. An inspection of~\eqref{eq:CHR_metric} reveals that the CHR bulk metric is constructed from the \enquote{seed} brane metric by generating exact copies of $g_{\mu\nu}$ at any given bulk \enquote{slice} with $y=$const., and then rescaling each one according to the value of the warp factor there $e^{2A(y)}$. Consequently, any brane curvature singularities, as well as any event horizons that are present in the seed metric, also extend indefinitely into the bulk, see left panel in Fig.~\ref{fig:CHR_vs_GEA_illustrated}.
On the other hand, the main idea behind the GEA is to perform the uplift of $g_{\mu\nu}$ in the bulk by generating copies of the metric in terms of a \emph{rotation} about the origin of the brane coordinates prior to warping the bulk spacetime, as is illustrated in the right panel of Fig.~\ref{fig:CHR_vs_GEA_illustrated}. This ensures that all features of $g_{\mu\nu}$ remain localized near the brane and do not~\enquote{leak} asymptotically into the bulk.

The starting point for the GEA, as with the CHR method, is the axisymmetric five-dimensional line element~\eqref{eq:Warped_flat_original}, with a $\bold{Z}_2$-symmetric bulk along the extra dimension, namely $A(y)=A(-y)$,\,\footnote{It is also possible to consider non-symmetric bulk braneworld models, however, in this case, the analysis for each part of the bulk should be performed separately.} and $A(0)=0$.
Unlike the CHR method however, in GEA, Eq.~\eqref{eq:Warped_flat_original} gets properly transformed prior to introducing any curvature on the brane. Upon rescaling $y$ according to
\beq
\dd y=e^{A(y)}\dd z \,\,\Rightarrow\,\, z(y)= \int_0^{y}\,e^{-A(\bar{y})}\,\dd \bar{y}\,,
\label{eq:GEA_z_def}
\eeq
with the new coordinate $z(y)$ being also $\bold{Z}_2$-symmetric, the brane located at $z=0$ and $z \in (-\infty,+\infty)$, the line element is brought to the form
\beq
\dd s^2=e^{2A(y(z))}\left(-\dd t^2+\dd r^2+\dd z^2+r^2\dd \Omega^2_2 \right)\,.
\label{eq:Flat_brane_r_z}
\eeq
The definition~\eqref{eq:GEA_z_def}, the AdS$_5$ asymptotics, and the condition $A(0)=0$, restrict the braneworld models that can be considered in the context of the GEA to those with injective and strictly monotonically decreasing warp function at each part of the bulk~\footnote{In the following sense: $A'(z)<0$ if $z>0$ and $A'(z)>0$ if $z<0$. This ensures that the warp function exhibits a global maximum at the location of the brane.} i.e. such that the domain of $z\in(-\infty,+\infty)$ has a one-to-one mapping with the domain of $y \in (-\infty,+\infty)$.

Subsequently, the algorithm calls for introducing $5$D spherical coordinates $\{\rho,\chi\}$~\cite{NK1}, by means of the invertible coordinate transformation
\ba
\{r,z\} &\to& \{\rho \sin \chi, \rho \cos\chi \}\,,\label{eq:rho_chi_def}\\
\{\rho,\chi \} &\to& \{\sqrt{r^2+z^2},\arctan\left(r/z\right) \}\,,\label{eq:rho_chi_inv}
\ea
with domains $\rho \in [0,+\infty)$ and $\chi\in [0,\pi]$. In this coordinate frame, the brane is located at $\chi=\pi/2$.\,\footnote{This is strictly true only for thin-brane models.
In thick-brane models, the brane can be thought of as a stack of timelike hypersurfaces around $\chi=\pi/2$. See \cite{Dzhunushaliev:2009va} for a review on thick-brane models and solutions.} Thus, we may rewrite the line element~\eqref{eq:Flat_brane_r_z} as
\beq
\dd s^2=e^{2A (y (\rho \cos\chi))}\left(-\dd t^2+\dd \rho^2+\rho^2\dd \Omega^2_3 \right)\,.
\label{eq:Flat_brane_rho_chi}
\eeq
The final step in the GEA consists of introducing curvature to the unwarped  flat metric $g_{MN}$ in~\eqref{eq:Flat_brane_rho_chi}, by promoting the $g_{00}$ and $g_{11}$ metric components to arbitrary functions of the bulk radial coordinate, leading to the five-dimensional geometry
\beq
\dd s^2=e^{2A(y(\rho \cos\chi))}\left(-f(\rho)\dd t^2+\frac{\dd \rho^2}{h(\rho)}+\rho^2\,\dd \Omega^2_3 \right)\,.
\label{eq:GEA_rho_chi}
\eeq
A very useful representation of the GEA metric~\eqref{eq:GEA_rho_chi} for the study of the five-dimensional structure of spacetime is obtained by following the inverse coordinate transformations~\eqref{eq:rho_chi_inv} resulting to
\beq
\dd s^2=e^{2A(y(z))}\left[-f(\rho)\dd t^2+\left(\frac{r^2}{h(\rho)}+z^2\right)\frac{\dd r^2}{\rho^2}+\left(\frac{z^2}{h(\rho)}+r^2\right)\frac{\dd z^2}{\rho^2}+\left(\frac{1-h(\rho)}{h(\rho)} \right)\frac{2 rz\,\dd r\dd z}{\rho^2}+r^2\,\dd\Omega^2_2 \right]\,.
\label{eq:g_tild_r_z}
\eeq
Taking one step further, via~\eqref{eq:rho_chi_def} we also have the representation in terms of $\{r,y\}$
\beq
\dd s^2=e^{2A(y)}\left[-f(\rho)\dd t^2+\left(\frac{r^2}{h(\rho)}+z^2\right)\frac{\dd r^2}{\rho^2}+r^2\,\dd \Omega^2_2+\frac{2 rz\,e^{-A(y)}\,\left(1-h(\rho) \right)}{h(\rho)\,\rho^2}\dd r\dd y \right]+\left(\frac{z^2}{h(\rho)}+r^2\right)\frac{\dd y^2}{\rho^2}\,,
\label{eq:g_tild_r_y}
\eeq
where it is understood that $\rho = \rho(r,y)= \sqrt{r^2+z^2(y)}$, and the functional form of $z=z(y)$ is specified for any desired braneworld model characterized by a warp function $A(y)$ in terms of Eq.~\eqref{eq:GEA_z_def}. By inspection of~\eqref{eq:g_tild_r_y}, it is immediately evident that the axisymmetry w.r.t. $y$ and spherical symmetry w.r.t. $r$ for any $y_0=\text{const.}$ slice in the bulk are (as of course they should) preserved by the GEA metric~\eqref{eq:GEA_rho_chi}.

With the above presented GEA,~\emph{any} static spherically-symmetric ${\rm 4D}$ brane geometry written in terms of the curvature radial coordinate with a line element given by
\beq
\dd s_4^2=-f(r) \dd t^2+\frac{\dd r^2}{h(r)}+r^2\dd \Omega^2_2\,,
\label{eq:brane_curved}
\eeq
can be immediately embedded in~\emph{any} ${\rm 5D}$ single-brane braneworld model characterized by a warp function $A(y)$, simply by substituting the functional form of $f(r)$ and $h(r)$ for the desired induced brane metric~\eqref{eq:brane_curved} into any of the Eqs.~\eqref{eq:GEA_rho_chi}-\eqref{eq:g_tild_r_y} after replacing $r \to \rho$. For example, the Schwarzschild black hole (BH) of mass parameter $M$ in~\eqref{eq:brane_curved} is described by
\beq
f(r)=h(r)=1-\frac{2M}{r}\,.
\eeq
Taking $r \to \rho =\sqrt {r^2+z^2}$ according to~\eqref{eq:rho_chi_inv}, the ${\rm 5D}$ braneworld extension of the Schwarzschild geometry for any warp factor, is given in terms of $\{t,\rho,\chi,\vartheta,\varphi\}$ by~\eqref{eq:GEA_rho_chi} with
\beq
f(\rho)=h(\rho)=1-\frac{2M}{\rho}\,,
\eeq
or equivalently in terms of $\{t,r,\vartheta,\varphi,y\}$ by~\eqref{eq:g_tild_r_y} with
\beq
f(\rho(r,y))=h(\rho(r,y))=1-\frac{2M}{\sqrt{r^2+z^2(y)}}\,.
\eeq

A very important general conclusion that can be drawn from the form of the GEA metric~\eqref{eq:GEA_rho_chi} is that, since for~\emph{any} asymptotically flat brane metric~\eqref{eq:brane_curved} the metric functions satisfy
\beq
\lim_{r \to \infty}f(r)=1\,,\quad\lim_{r \to \infty}h(r)=1\,,
\eeq
the corresponding GEA metric functions will also satisfy
\beq
\lim_{\rho \to \infty}f(\rho)=1\,,\quad\lim_{\rho \to \infty}h(\rho)=1\,,
\eeq
and as such, the GEA line element~\eqref{eq:GEA_rho_chi} reduces to AdS$_5$ asymptotically into the bulk with a curvature radius that is determined by the warp factor. Even in the case of asymptotically (A)dS metrics on the brane, it has been demonstrated that the AdS$_5$ asymptotics of the GEA are guaranteed \cite{Nakas:2023yhj}. In this case, the~\emph{effective} asymptotic AdS$_5$ curvature radius is determined in terms of both the warp factor of the braneworld model under consideration, and the effective brane cosmological constant. Of course, in such cases, consistency relations must be imposed between the aforementioned parameters so that the effective bulk cosmological constant is negative~\cite{Nakas:2023yhj}. Finally, a direct byproduct of this observation is that the curvature invariants for any GEA metric will deviate from those of pure AdS$_5$ only near the brane. It is in this sense, that the bulk extensions provided by the GEA are guaranteed by construction to be brane-localized.

Let us now derive in direct analogy to the completely general analysis of Sec.~\ref{Sec:2}, the conditions on the functions of the GEA metric~\eqref{eq:GEA_rho_chi}, in order for it to represent a five-dimensional wormhole structure connecting two asymptotic regions of spacetime in single-brane braneworld models with injective (one-to-one) and non-singular warp factors. Since the original formulation of the GEA \cite{Nakas:2023yhj} is based on the curvature radial coordinate $r \geqslant 0$, it can be used to uplift~\emph{any} MT brane geometry~\eqref{eq:brane_curved} into the bulk~\eqref{eq:GEA_rho_chi}~\emph{including} four-dimensional WHs. However, restricting the domain of $r\geqslant R_{\Join}$ in order to describe a wormhole spacetime, cannot be done naively and a careful consideration of the domains of all the coordinates is necessary for a proper interpretation of the 5D structure of these objects. It is also important to recall (see discussion in Sec.~\ref{Sec:2}) that, the MT representation of a WH requires in general two sets of metric functions for the description of the entire spacetime when the wormhole is throat-asymmetric. Here, we perform the analysis on one side of the WH, where the observer would be located, and it is understood that for throat-asymmetric WHs an identical analysis would have to also be performed for the other side of the throat. In any case, the study of throat-asymmetric WHs is more naturally accommodated in the nMT frame which is discussed in Sec.~\ref{Sec:eGEA}.

For the definition of a wormhole structure in braneworld scenarios (in complete analogy to the general analysis of Sec.~\ref{Sec:2}), we begin with the requirement for the existence of a minimal-area spacelike ($t=\text{const.}$) hypersurface at a fixed value of the radial coordinate $\rho=\rho_0$. The line element of such a hypersurface in the GEA spacetime~\eqref{eq:GEA_rho_chi} becomes
\beq
\dd s_3^2 =e^{2A(y(\rho_0 \cos\chi))}\rho_0^2\dd \Omega_3^2\,,
\label{eq:GEA_spacelike_surface}
\eeq
and thus, we immediately identify the proper circumferential radius in the bulk as
\beq
R_{\rm eff}(\rho_0,\chi)\equiv e^{A(y(\rho_0 \cos\chi))}\rho_0\,.
\label{eq:R_MT_GEA_generic}
\eeq
Notice that the $\chi$-dependence in~\eqref{eq:R_MT_GEA_generic} is a manifestation of the underlying bulk axi-symmetry w.r.t. the extra dimension. Note also that for throat-symmetric WHs, where the MT representation of the spacetime is isometric on both sides of the throat, the functional form of~$R_{\rm eff}$ is solely determined by the warp factor of the braneworld model under consideration, and the physical radius of the seed brane metric. It is thus completely independent from the details of the metric functions of the brane seed-metric. In the case of throat-asymmetric WHs, such as e.g.~the Ellis-Bronnikov or~\enquote{anti-Fisher} solution studied in Sec.~\ref{Sec:EB}, the non-trivial profile of the metric function $\rho(u)$, which in general can be non-invertible, induces a dependence of $R_{\rm eff}$ on the details of the seed brane metric.
In the brane limit, $R_{\rm eff}(\rho_0,\pi/2)$ reduces to the four-dimensional circumferential radius $r_0$.

An important observation has to be made as this point.~The functional form of the bulk circumferential radius~\eqref{eq:R_MT_GEA_generic} has a non-trivial dependence on the bulk radial coordinate $\rho$ in contrast to the circumferential radius metric function of the MT seed brane metric that generated it. Naively then, one may be tempted to determine the coordinate location of the hyperthroat in the bulk by following the nMT minimization conditions~\eqref{eq:r0_cond_1},\,\eqref{eq:r0_cond_2}. Let us demonstrate why this is not the correct approach for braneworld MT WHs. First we perform, in analogy with~\eqref{eq:total_area}, the spacelike hypersurface integral at constant $\rho_{\rm 0}$ for~\eqref{eq:GEA_spacelike_surface} in order to obtain the total surface area
\beq
S(\rho_0)=4\pi \int_{0}^{\pi}R_{\rm eff}^3(\rho_0,\chi) \sin^2\chi\, \dd\chi\,.
\label{eq:GEA_total_area}
\eeq
Using~\eqref{eq:R_MT_GEA_generic} and~\eqref{eq:rho_chi_def}, the above integral can be written equivalently in terms of the bulk coordinate $z$ as follows
\beq
S(\rho_0)=4\pi \int_{-\rho_0}^{\rho_0}e^{3A(y(z))}\rho_0\sqrt{\rho_0^2-z^2}\, \dd z\,.
\label{eq:GEA_total_area_dz}
\eeq
In the weak bulk warping limit where $e^{A}\simeq 1$, the integral~\eqref{eq:GEA_total_area_dz} reduces to $S(\rho_0)\simeq 2\pi^2\rho_0^3$, i.e. to the total area of a sphere of radius $\rho_0$ in four-dimensional flat space. This limit is realized when $\rho_0$ is much smaller than the curvature radius of the AdS$_5$ bulk. Consequently, for any constant-$\rho_0$ hypersurface, the warping of the extra dimension, as encoded in the presence of a non-trivial warp factor in~\eqref{eq:GEA_total_area_dz}, reduces the total surface area.
The surface area extremization condition corresponds to the vanishing of the first $\rho_0$-derivative of~\eqref{eq:GEA_total_area_dz} that upon utilizing the Leibniz integral rule reads 
\beq
\frac{\dd S(\rho_0)}{\dd \rho_0}=4\pi \int_{-\rho_0}^{\rho_0}e^{3A(y(z))}\frac{2\rho_0^2-z^2}{\sqrt{\rho_0^2-z^2}}\, \dd z\,,
\label{eq:MT_extr_cond_GEA}
\eeq
where at $z\to\pm\rho_0$ the boundary terms vanish and the integral converges. For all asymptotically AdS$_5$ braneworld models, with $A(0)=0$ and an injective and strictly monotonically decreasing warp function at each part of the bulk, the integral in \eqref{eq:MT_extr_cond_GEA} is always strictly positive and finite leading to the general conclusion
\beq
 0<\frac{\dd S(\rho_0)}{\dd \rho_0}<+\infty\,,\quad\forall\, \rho_0<+\infty\,.
\label{eq:dSdrho_positivity}
\eeq
As a result, the extremization condition has no solutions for $\rho_0>0$. The interpretation of~\eqref{eq:dSdrho_positivity} is that the total surface area of the warped spheres of constant radial coordinate $\rho_0$ is a monotonically increasing function of $\rho_0$ in direct analogy with the case of flat
extra dimensions.

We can thus conclude that one cannot naively apply the nMT conditions in this case despite the non-trivial bulk circumferential radius. The proper way to determine $\rho_{\rm th}$ then comes, as with any MT WH in flat-dimensions, from the divergence of the $g_{11}$ in~\eqref{eq:GEA_rho_chi} corresponding to
\beq
e^{-2A(y(\rho_{\rm th} \cos\chi))}h(\rho_{\rm th})=0\quad\Rightarrow\quad h(\rho_{\rm th})=0\,.
\label{eq:Throat_location_GEA_MT}
\eeq
In the last step, we used the assumption that for any reasonable braneworld model, the warp factor is non-vanishing everywhere in the bulk. We have thus discovered that for MT WHs, the condition for the determination of the coordinate location of the throat is the same in any number of flat-extra dimensions and in 5D single brane braneworld models with a non-singular and injective warp factor. The physical radius of the wormhole hyperthroat is then given by~\eqref{eq:R_MT_GEA_generic} evaluated at the solution $\rho_{\rm th}$ of~\eqref{eq:Throat_location_GEA_MT} and it reduces to the four-dimensional value on the brane.

Having determined the coordinate location of the hyperthroat, we are now in position to discuss the domains of the various coordinates. The restricted domain of the radial coordinate $r\geqslant R_{\Join}$ for the seed brane wormhole metric, induces a lower cut-off on the bulk radial coordinate $\rho$ as defined in~\eqref{eq:rho_chi_inv} in the following way
\beq
\rho\equiv\sqrt{r^2+z^2}\,\Rightarrow\,\rho\geqslant R_{\Join}\,,
\label{eq:R_throat_condition_MT}
\eeq
given that $z \in (-\infty,+\infty)$. This means that, $R_{\Join}$ corresponds to the coordinate location of the hyperthroat i.e. $\rho_{\rm th}=R_{\Join}$, and this is indeed the same solution one would get from~\eqref{eq:Throat_location_GEA_MT}, having used the functional form of the 4D seed metric function: $h(r)$ with $r\geqslant R_{\Join}$. Consequently, Eq.~\eqref{eq:R_throat_condition_MT} implies the following constraint between $r$ and $z$ at the location of the hyperthroat
\beq
R_{\Join}^2=r^2+z^2\,,
\label{eq:rho0_constraint}
\eeq
defining also the boundary of spacetime, see also Fig.~\ref{Fig: coords}.

\begin{figure*}
\centering
\begin{tikzpicture}[auto, scale=1, every node/.style={scale=1}]

\definecolor{mypurple}{HTML}{e82beb80}


\draw[-,ultra thick, dashed] (0,0)--(1.4,1.4);
\draw[-,ultra thick] (0,0)--(-1.4,1.4) node[midway, above]{$R_{\Join}$};

\draw[fill=mypurple, fill opacity=0.5] ([shift=(0:2)] 0,0) arc (0:180:2);

\draw[->,ultra thick] (0,0)--(0,4) node[above]{$r$};
\draw[->,ultra thick] (-4,0)--(4,0) node[right]{$z$};

\draw[->,ultra thick] (1.4,1.4)--(2.5,2.5) node[above]{$\rho$};

\draw [ultra thick]  ([shift=(0:0.6)] 0,0) arc (0:45:0.6) node[midway, right]{$\chi$};

\end{tikzpicture}
\caption{Coordinates in the bulk. The light-purple region is not part of spacetime, since the radial coordinate $\rho\geqslant R_\Join$. However this does not restrict the domains for the coordinates $r \in [0,+\infty)$ and $z\in (-\infty,+\infty)$.}
\label{Fig: coords}
\end{figure*}

Equation~\eqref{eq:rho0_constraint} reveals that the domain of $r$ is not globally $r\geqslant R_{\Join}$, an assumption that would have led to a misinterpretation of the 5D wormhole structure, but it depends on the value of $z$. On the brane $(z=0)$ we have $r \in [R_{\Join},+\infty)$, thus recovering the 4D domain of the seed metric, while, in the bulk, the lower cutoff is reduced as we move away from the brane and further into the bulk according to
\beq
r_{\rm min}(z)=\sqrt{R_{\Join}^2-z^2}\,,\quad \forall\, z \in [-R_{\Join},R_{\Join}]\,.
\label{eq:r_min_domain}
\eeq
Consequently, after the uplift of the 4D WH in a 5D braneworld model, the domain of the four-dimensional radial coordinate is expanded: $r \in [0,+\infty)$. 
Conversely, the original domain of the $r$ coordinate is inherited in the five-dimensional radial coordinate, namely $\rho\in [R_\Join,+\infty)$. 

The next criterion for a wormhole structure in the MT frame are the flare-out conditions. We begin by considering $z=z_{\rm 0}=\text{const.}$ slices in the $\{r,z\}$ representation of the GEA metric~\eqref{eq:g_tild_r_z} for which the induced spacelike hypersurface ($t=\text{const.}$) with $\vartheta=\pi/2$ is described by the following line element 
\beq
\dd s_2^2=e^{2A(y(z_{\rm 0}))}\left(\frac{r^2}{h\left(\sqrt{r^2+z_{\rm 0}^2}\right)}+z_{\rm 0}^2 \right)\frac{\dd r^2}{r^2+z_{\rm 0}^2}+e^{2A(y(z_{\rm 0}))} r^2\dd \varphi^2\,.
\label{eq:spacelike_z0}
\eeq
Then, by utilizing the spherical symmetry w.r.t. $r$ for each $z_0$ slice in the bulk we are able to apply the standard methodology~\cite{Kim:2013tsa}. A rescaling of the radial coordinate according to
\beq
r \to x\equiv e^{A(y(z_{\rm 0}))} r\,,
\eeq
recasts~\eqref{eq:spacelike_z0} in the form
\begin{equation}
    \dd s_2^2=\frac{\dd x^2}{H(x)}+x^2 \dd \varphi^2\,,
    \label{eq:Hx_def}
\end{equation}
with
\begin{equation}
    H(x)\equiv \big(e^{-2A(y(z_{\rm 0}))}x^2+z_{\rm 0}^2\big)\left(\frac{ e^{-2A(y(z_{\rm 0}))}x^2}{h\left(\sqrt{ e^{-2A(y(z_{\rm 0}))}x^2+z_{\rm 0}^2}\right)}+z_{\rm 0}^2 \right)^{-1}\,.
\end{equation}
Next, we consider the embedding three-dimensional Euclidean space \footnote{The term "embedding space" refers to a higher-dimensional space in which another brane or surface is embedded.} in cylindrical coordinates $\{l,x,\varphi\}$: 
\begin{equation}
    \dd s_3^2=\dd l^2+\dd x^2+x^2\dd \varphi^2\,.
\end{equation}
The variable $x$, corresponding to the radius in cylindrical coordinates, has to be non-negative, however, $l$ has no such restriction and its domain is $l \in \mathbb{R}$. Upon allowing $l$ to depend on $x$, the above space becomes
\beq
\dd s_2^2=\left[ 1+\left(\frac{\dd l}{\dd x}\right)^2\right] \dd x^2+x^2\dd \varphi^2\,.
\label{eq:Aux_Euclidean_final}
\eeq
By identifying~\eqref{eq:Aux_Euclidean_final} with~\eqref{eq:Hx_def}, we obtain the equation that determines the generalized~\emph{lift} function $l(x)$
\beq
\frac{\dd l(x)}{\dd x}=\pm\sqrt{\frac{1-H(x)}{H(x)}}\,,
\label{eq:lift_func_x}
\eeq
and thus, the flare-out condition at the throat is given by
\beq
\frac{d^2x}{dl^2}=\frac{H'(x)}{2(H(x)-1)^2}>0\,.
\label{eq:flare_out_x}
\eeq
Equations~\eqref{eq:lift_func_x} and~\eqref{eq:flare_out_x} can be expressed in terms of $r$ at each $z_0=\text{const.}$ slice in the bulk respectively by\,\footnote{Rigorously, a different symbol should be used for $\widetilde{l}(r)\equiv l(x(r))$ but we choose to avoid introducing another symbol here.}
\beq
l(r)=\pm\int_{r_{\rm 0}}^{r} e^{A(y(z_0))} \sqrt{\frac{\left[1-h\left(\sqrt{\bar{r}^2+z_0^2}\right)\right]\bar{r}^2}{h\left(\sqrt{\bar{r}^2+z_0^2}\right)\left(\bar{r}^2+z_0^2 \right)}}~ \dd\bar{r}\,,
\label{eq:lift_func_r}
\eeq
and
\beq
\frac{ e^{-A(y(z_0))}\left\{ 2z_0^2\,h\left(\sqrt{r^2+z_0^2}\right) \left[ h\left(\sqrt{r^2+z_0^2}\right)-1\right]+r\left(r^2+z_0^2\right) \partial_{r}h\left(\sqrt{r^2+z_0^2}\right) \right\}}{2 r^3 \left[ h\left(\sqrt{r^2+z_0^2}\right) -1 \right]^2}>0\,.
\label{eq:flare_out_r}
\eeq
Notice that on the brane $(z_0=0)$,~Eqs.~\eqref{eq:lift_func_r} and~\eqref{eq:flare_out_r} reduce to the standard four-dimensional expressions~\cite{Kim:2013tsa}. Furthermore, as we have discussed in Sec.~\ref{Sec:2}, the four-dimensional expressions have the same functional form for any number of flat extra dimensions. Thus, we have discovered that the warping of the extra dimensions, modifies the flare out conditions for higher-dimensional WHs in a non-trivial way.
The last condition for the GEA metric~\eqref{eq:GEA_rho_chi} to describe a two-way traversable WH, is the absence of event horizons and this corresponds to $f(\rho)>0$ and $h(\rho)\geqslant0$ everywhere.

\subsection{The extended formulation of the GEA and non-Morris-Thorne wormholes}
\label{Sec:eGEA}

\begin{figure*}[h]
\centering
\begin{tikzpicture}[auto, scale=0.95, every node/.style={scale=0.95}]

\tikzstyle{place0}=[rectangle,draw=white,fill=white]
\tikzstyle{place1}=[rectangle,rounded corners,draw=black!50,fill=orange!40,thick]
\tikzstyle{place2}=[rectangle,draw=black!,fill=white,rounded corners,thick]
\tikzstyle{place3}=[rectangle,rounded corners,draw=black!50,fill=gray!20,thick]
\tikzstyle{pre}=[<-,shorten <=1pt,>=stealth’,semithick]
\tikzstyle{post}=[->,shorten >=1pt,>=stealth’,semithick]

\node[align=center] (5Dbulk) at (0,+3) [place0] {\textbf{\large{5D bulk}}\\[0.1mm] \textbf{spacetime}};

\node[align=center] (brane) at (0,-1.5) [place0] {\textbf{\large{4D brane}}};

\draw[black,line width=0.5mm,dashed] (-0.5,0) -- (17.5,0);

\draw[black,line width=0.5mm,dashed] (1,5.5) -- (1,-2.5);

\node[align=center] (5DMT) at (6,5) [place1] {Localized braneworld WH\\[0.1mm] in MT frame. Equation \eqref{eq:GEA_rho_chi}};
\node[inner sep=2pt] (throatMT) at (3.5,2) [place2] 
{\begin{minipage}{0.2\linewidth}
    \centering
    \includegraphics[width=.4\textwidth]{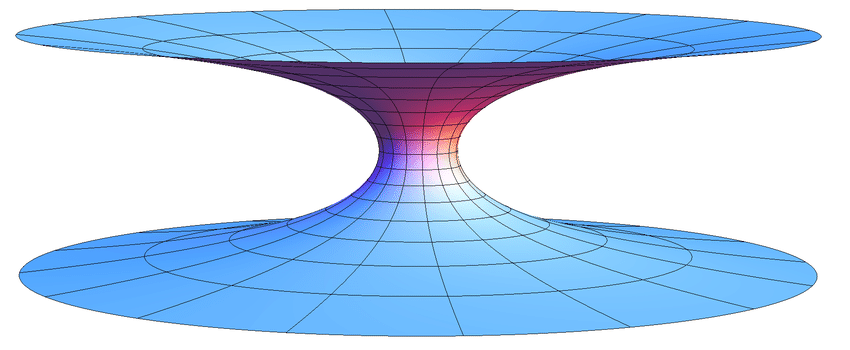} \\
    Wormhole throat determined by the flare-out condition \eqref{eq:flare_out_r} and \eqref{eq:Throat_location_GEA_MT}
    \end{minipage}};

\node[align=center] (5DnMT) at (13,5) [place1] {Localized braneworld WH\\[0.1mm] in nMT frame. Equation \eqref{eq:ds2_eGEA}};
\node[inner sep=2pt] (throatnMT) at (15.5,2) [place2] 
{\begin{minipage}{0.2\linewidth}
    \centering
    \includegraphics[width=.4\textwidth]{throat.png} \\
    Wormhole throat determined by the minimization of the total area. Equations \eqref{eq:nMT_extr_cond2} and \eqref{eq:nMT_min_con}
    \end{minipage}};

\node[align=center] (4DMT) at (6,-2) [place3] {Wormhole in MT frame.\\[0.1mm]
Equation \eqref{eq:brane_curved}};

\node[align=center] (4DnMT) at (13,-2) [place3] {Wormhole in nMT frame.\\[0.1mm]
Equation \eqref{eq:ds2_nMT_brane}};

\draw[very thick,black,<->] (5DMT.south) .. controls +(0,-1) and +(0,1) .. (4DMT.north) node[draw=none,fill=none,font=\scriptsize,midway,right] {\large{\textbf{GEA}}};
\draw[thick,black,->] (5DMT.west) .. controls +(-1,0) and +(0,1) .. (throatMT.north) node[draw=none,fill=none,font=\scriptsize,midway,right] {};
\draw[thick,black,->] (throatMT.south) .. controls +(0,0) and +(-1,0) .. (4DMT.west) node[draw=none,fill=none,font=\scriptsize,midway,right] {$z_0=0$};

\draw[very thick,black,<->] (5DnMT.south) .. controls +(0,-1) and +(0,1) .. (4DnMT.north) node[draw=none,fill=none,font=\scriptsize,midway,left] {\large{\textbf{eGEA}}};
\draw[thick,black,->] (5DnMT.east) .. controls +(1,0) and +(0,1) .. (throatnMT.north) node[draw=none,fill=none,font=\scriptsize,midway,right] {};
\draw[thick,black,->] (throatnMT.south) .. controls +(0,0) and +(1,0) .. (4DnMT.east) node[draw=none,fill=none,font=\scriptsize,midway,right] {$\chi=\pi/2$};

\draw[very thick,black,->] (5DMT.east) .. controls +(1,0) and +(-1,0) .. (5DnMT.west) node[draw=none,fill=none,font=\scriptsize,midway,above] {$\rho=\rho(u)$};
\draw[very thick,black,->] (4DMT.east) .. controls +(1,0) and +(-1,0) .. (4DnMT.west) node[draw=none,fill=none,font=\scriptsize,midway,above] {$r=r(u)$};

\end{tikzpicture}

\vspace{1em}
\caption{Schematic representations of both the General Embedding Algorithm (GEA) for the Morris-Thorne (MT) frame and the extended General Embedding Algorithm (eGEA) for the non-Morris-Thorne (nMT) frame.}
\label{Fig: algorithm}
\end{figure*}

With the GEA as presented in the previous section, it is possible to embed four-dimensional spacetimes into the bulk assuming that the brane metric is written in the MT frame where the radial coordinate is the curvature coordinate. In principle, however, one may start with a four-dimensional metric in terms of an arbitrary radial coordinate $u \in (-\infty,+\infty)$ (nMT frame)
\beq
\dd s_4^2=-\widetilde{f}(u)\dd t^2+\frac{\dd u^2}{\widetilde{h}(u)}+r^2(u) d\Omega^2_{2}\,.
\label{eq:ds2_nMT_brane}
\eeq
Such a line element is commonly used for the representation of wormhole spacetimes where the circumferential radius metric function $r(u)$ has the restricted domain $r \in [R_{\Join},+\infty)$ with the physical radius of the throat located at coordinate value $u_{\rm th}$ and given by $r(u_{\rm th})=R_{\Join}>0$. For~\emph{invertible} $r(u)$, the radial coordinate transformation $\dd r^2=\dd u^2 \left[r'(u)\right]^2$, recasts~\eqref{eq:ds2_nMT_brane} into the MT frame
\ba
\dd s_4^2&=&-\widetilde{f}(u(r))\dd t^2+\frac{\dd r^2}{\widetilde{h}(u(r))\left[r'(u(r))\right]^2}+r^2 d\Omega^2_{2}
\label{eq:ds2_nMT_brane_intermediate}\\
&=& -f(r) \dd t^2+\frac{\dd r^2}{h(r)}+r^2\dd \Omega^2_2\,,
\label{eq:MT_4D_eGEA}
\ea
where $f(r) \equiv \widetilde{f}(u(r))$ and $h(r) \equiv \left[r'(u(r))\right]^2\widetilde{h}(u(r))$. Then, one is able to indirectly uplift the metric~\eqref{eq:ds2_nMT_brane}, via~\eqref{eq:MT_4D_eGEA} in any single-brane braneworld scenario by means of the GEA. Thus, it becomes evident that the circumferential radius metric function $r(u)$, corresponding to the natural choice of a radial coordinate, plays an important role in the uplift of any 4D metric, independently of the radial coordinate used to represent the brane metric. Nevertheless, this indirect approach to the uplift that is based on intermediate MT frames, fails when $r(u)$ is~\emph{non-invertible} and a new, extended formulation of the GEA (eGEA) becomes necessary. See also~Fig.~\ref{Fig: algorithm}, where a schematic representation of the relation between the GEA and the eGEA as methods for the uplift of brane seed metrics in various coordinate systems is presented.

The extended formulation of the GEA, emerges quite naturally if we use one of the key observations from the previous section on the uplift of MT wormholes. As we have shown, the proper embedding of branes with non-trivial topology requires that the restricted domain of the brane radial coordinate $r$ has to be inherited by the bulk radial coordinate $\rho$. This effectively corresponds to a~\enquote{top-down} or~\enquote{bulk-first} approach to the embedding of 4D spacetimes. Building on this insight, we follow the exact same steps as in the original formulation of the GEA leading up to~\eqref{eq:GEA_rho_chi}. Then, we promote the bulk radial coordinate $\rho \geqslant 0$ to a function of an arbitrary radial coordinate,~$\rho\to\rho(u)$ with $u \in (-\infty,+\infty)$ and $\rho(u_{\rm th})=R_{\Join}>0$, thus recasting~\eqref{eq:GEA_rho_chi} to
\beq
\dd s^2=e^{2A(y(\rho(u) \cos\chi))}\left[ -f\left( \rho(u)\right)\dd t^2+\frac{\left( \rho'(u)\right) \dd u^2}{h\left(\rho(u)\right)}+\rho^2(u)\dd \Omega^2_3\right]\,.
\eeq
After a redefinition of the metric functions $g_{00}$ and $g_{11}$ we have the final form for the extended formulation of the GEA given by
\beq
\dd s^2=e^{2A(y(\rho(u) \cos\chi))}\left[-\widetilde{f}(u) \dd t^2+\frac{\dd u^2}{\widetilde{h}(u)}+\rho^2(u) \dd \Omega^2_3\right]\,.
\label{eq:ds2_eGEA}
\eeq
As a consequence of the non-trivial form of $\rho(u)$, the coordinate transformation~\eqref{eq:rho_chi_def} now becomes
\ba
\label{eq:nMT-coord-trans}
\{r\left(u,\chi\right),z\left(u,\chi\right)\} &\to& \{\rho(u) \sin \chi, \rho(u) \cos\chi \}\,,\\
\{\rho(u),\chi \} &\to& \{\sqrt{r^2+z^2},\tan^{-1}\left(r/z\right) \}\,,
\ea
In the brane limit of $\chi \to \pi/2$ then, one obtains
\beq
r(u,\pi/2)=\rho(u)\,,
\label{eq:r_rho_eGEA_bridge}
\eeq
and thus the bulk radial coordinate smoothly transitions to the brane one. Equation~\eqref{eq:r_rho_eGEA_bridge} then, corresponds to the key equation for the uplift of~\eqref{eq:ds2_nMT_brane} directly into the eGEA metric~\eqref{eq:ds2_eGEA} for~\emph{any} function $r(u)$,~\emph{without any requirement for invertibility}, and for any warp function $A(y)$ that respects the same criteria as for GEA. In alignment with the original GEA, this is achieved simply by substituting the metric functions from~\eqref{eq:ds2_nMT_brane} into~\eqref{eq:ds2_eGEA}, having promoted $r(u)\to\rho(u)$, and ensuring that bulk radial coordinate appearing in the warp factor has the same functional form i.e. $\rho(u)$. We provide examples for the application of the eGEA in sections~\ref{Sec:SV} and~\ref{Sec:EB}.

It should be noted that, by construction of the line element~\eqref{eq:ds2_eGEA}, the presence or absence of $\bold{Z}_2$ symmetry in the bulk remains a property of the 5D spacetime that is solely determined by the functional form of the warp function and~\emph{it is independent} of the symmetry or asymmetry of $\rho(u)$ with respect to the location of the WH throat. Furthermore, in analogy with the arguments presented for the original GEA metric, we remark that the localization of the five-dimensional line element~\eqref{eq:ds2_eGEA} is once again guaranteed for any asymptotically-flat seed brane metric. The necessary and sufficient conditions for asymptotic flatness of the seed brane metric in terms of an arbitrary radial coordinate $u$~\cite{Bronnikov:2012wsj,Bronnikov:2021liv}, result to the following conditions on the eGEA metric functions that guarantee localization 
\begin{equation}
\lim_{|u|\to +\infty}\widetilde{f}(u) ={\rm const.}\,,\quad \lim_{|u|\to +\infty}\widetilde{h}(u)\left(\frac{\dd \rho}{\dd u} \right)^2 = 1\,.
\end{equation}
It is also important to emphasize that the relevance of the eGEA as an uplift method, extends beyond the applications for wormhole spacetimes, since there are examples of black-hole geometries (e.g.~\cite{Holzhey:1991bx}) which are obtained in the nMT frame in terms of a non-invertible $r(u)$. The uplift of such metrics into braneworld models requires the eGEA, and the study of their properties presents an interesting direction for future research.

Pertaining now to wormhole structures in~\eqref{eq:ds2_eGEA}, we follow the same analysis as in Secs. \ref{Sec:2} and \ref{Sec:MT}.
The line element for a  spacelike ($t=$~const.) hypersurface in the bulk spacetime \eqref{eq:ds2_eGEA} at a fixed value of the radial coordinate $u=u_0$ reads
\begin{equation}
ds_3^2=R^2_{\rm eff}(u_0,\chi)\,\dd \Omega_3^2\,,
\label{eq:nMT-hyper}
\end{equation}
with the circumferential radius being given by
\begin{equation}
\label{eq:Reff-nMT}
R_{\rm eff}(u_0,\chi)=e^{A(y(\rho(u_0) \cos\chi))}\rho(u_0)\,.
\end{equation}
In complete analogy with equation~\eqref{eq:GEA_total_area}, the total surface-area of the hypersurface \eqref{eq:nMT-hyper} is determined by the integral
\begin{equation}
S(u_0)=4\pi \int_{0}^{\pi}R_{\rm eff}^3(u_0,\chi) \sin^2\chi\, \dd\chi\,,
\label{eq:S_u0}
\end{equation}
and by using \eqref{eq:nMT-coord-trans} it can be expressed in terms of the $z$-coordinate as follows
\begin{equation}
    S(u_0)=4\pi \int_{-\rho(u_0)}^{\rho(u_0)}e^{3A(y(z))}\rho(u_0)\sqrt{\rho^2(u_0)-z^2}\, \dd z\,.
\end{equation}
The extremization condition here can be written as
\beq
\label{eq:nMT_extr_cond1}
\frac{\dd S(u_0)}{\dd u_0} = \mathcal{I}(u_0) \frac{\dd \rho (u_0)}{\dd u_0}=0\,,
\eeq
where the integral $\mathcal{I}(u_0)$ is given by 
\begin{equation}
\mathcal{I}(u_0) = 4\pi \int_{-\rho(u_0)}^{\rho(u_0)}e^{3A(y(z))}\frac{2\rho^2(u_0)-z^2}{\sqrt{\rho^2(u_0)-z^2}}\, \dd z\,.
\end{equation}
For any warp factor with the necessary characteristics allowed by GEA and eGEA (see below~\eqref{eq:Flat_brane_r_z}), the function $\mathcal{I}(u_0)$ is non-negative, smooth and monotonically increasing $\forall\, u_0\in(-\infty,+\infty)$. 
Consequently, the surface-area extremization condition for nMT braneworld WHs reduces to that of nMT wormholes in flat-extra dimensions namely
\beq
\label{eq:nMT_extr_cond2}
\frac{\dd S(u_0)}{\dd u_0}=\frac{\dd \rho (u_0)}{\dd u_0}=0\,.
\eeq

The minimization condition requires the positivity of the second derivative of the total area w.r.t.~$u_0$.
By using \,\eqref{eq:nMT_extr_cond1} it is straightforward to deduce that this condition leads to the inequality
\begin{equation}
    \label{eq:nMT_min_cond}
    \frac{\dd^2 S(u_0)}{\dd u_0^2}=\frac{\dd \mathcal{I}(u_0)}{\dd u_0}\frac{\dd \rho(u_0)}{\dd u_0}+\mathcal{I}(u_0)\frac{\dd^2 \rho(u_0)}{\dd u_0^2}>0\,.
\end{equation}
At first glance, the derivative $\dd\mathcal{I}/\dd u_0$ seems indeterminate, however, it can be shown that for all $u_0$ its value is finite.
Notice that for any allowed warp factor, $\mathcal{I}(u_0)$ is bounded from both below and above according to the following inequality:
\begin{equation}
    0\leq \mathcal{I}(u_0)\leq 4\pi \int_{-\rho(u_0)}^{\rho(u_0)}\frac{2\rho^2(u_0)-z^2}{\sqrt{\rho^2(u_0)-z^2}}\, \dd z =6\pi^2\rho^2(u_0)\,.    
 \end{equation}
Since $\mathcal{I}(u_0)$ is a smooth monotonically increasing function with a finite upper bound, its derivative at any $u_0$ will be finite as well.
As a result, the first term in \eqref{eq:nMT_min_cond} vanishes at the extremum, and the minimization condition is expressed as
\begin{equation}
    \label{eq:nMT_min_con}
    \frac{\dd^2 S(u_0)}{\dd u_0^2}=\frac{\dd^2 \rho(u_0)}{\dd u_0^2}>0\,.
\end{equation}
Overall, we have found that the minimization conditions defining the coordinate location of the throat of braneworld WHs are identical to those in the case of flat extra dimensions.
The above findings imply that the extrema of the circumferential radius in the uplifted geometry retain their classifications as throats, or equators if present, as in cases of seed brane wormholes with multiple throats~\cite{Antoniou:2019awm}.

\subsection{Stress-energy tensor and energy conditions}
\label{Sec:En-conds}

For any braneworld wormhole that one could consider using either the GEA or eGEA, the investigation of the energy conditions is of essential importance. 
The study of the energy conditions, in the causal structure of spacetime, reveals whether the matter/energy that is required to support the spacetime geometry respects some basic laws of nature.
However, depending on the form of the field theory, it is not always clear how to interpret the stress-energy tensor, especially in theories in which matter fields are coupled with gravitational terms.
Characteristic examples are scalar-tensor theories with non-minimal couplings with gravity, e.g. Einstein-scalar-Gauss-Bonnet models, Horndeski gravity \cite{Horndeski:1974wa}, Chern-Simons gravitational theories, and so on.
In all of these scenarios, matter can be coupled to gravity in multiple ways, either with the Ricci scalar (Einstein-Hilbert term), the Einstein tensor (Horndeski gravity), the Gauss-Bonnet term, or the Chern-Simons term (see e.g. \cite{Antoniou:2019awm, Chatzifotis:2020oqr, Bakopoulos:2021liw, Chatzifotis:2021hpg, Karakasis:2021tqx, Chatzifotis:2022mob, Bakopoulos:2023tso} and references therein).
Contrary to GR, in such theories, it is, more often than not, impossible to separate gravity from the matter fields at the level of equations of motion.
By forcing the equations of motion to be written in the standard form $G_{\mu\nu}=T_{\mu\nu}$, the interpretation of gravitational terms as part of the energy-momentum tensor of matter becomes unavoidable.
The way that one interprets the stress-energy tensor can utterly affect the conclusions drawn regarding the energy conditions.
In the literature, there are different interpretations regarding the allowed interactions that can be incorporated in $T_{\mu\nu}$, and in some cases, the energy-momentum tensor supporting a wormhole configuration may satisfy all of the energy conditions \cite{Kar:2015lma, Antoniou:2019awm}.

The analysis of the energy conditions, in this article, will be based on the assumption that the action functional leading to the braneworld WHs is of the form
\begin{equation}
    \label{eq: act}
    S=\frac{1}{2\kappa_5}\int d^5x \sqrt{-g}\left(R + \mathcal{L}_M \right)\,,
\end{equation}
where gravity and matter are decoupled.
In geometrized units ($c=G=1$), $\kap_5$ has dimensions of $(\text{length})$, while both the Ricci scalar and the matter Lagrangian density are on equal footing with dimensions of $(\text{length})^{-2}$.
The action \eqref{eq: act} leads to the well-known Einstein equations
$G_{MN}=T_{MN}$.
In this category of theories, one is able to use the Einstein tensor instead of the energy-momentum tensor to define all the necessary quantities and study the energy conditions.
We know from previous studies, regarding localized braneworld black holes in the context of the RS-II model~\cite{NK1,NK2,Neves:2021dqx}, that the exponential warping of the bulk spacetime results in a violation of the energy conditions away from the brane.
This happens because the AdS$_5$ asymptotic nature of the 5D spacetime rapidly dominates in the bulk, thus rendering all the energy conditions to be violated.
However, on and in the vicinity of the brane it has been shown that the energy conditions can be satisfied.
In what follows, we consider the metric \eqref{eq:ds2_eGEA}, which represents the braneworld-uplifted version of a general four-dimensional, spherically symmetric spacetime expressed in the most general coordinate system, i.e. the nMT frame. 
Our focus will be on examining the essential properties of the stress-energy tensor required to support this geometry, as well as studying the associated energy conditions.
The $T^{MN}$ necessary to support the spacetime geometry \eqref{eq:ds2_eGEA}, under the assumption \eqref{eq: act}, is of the form
\begin{equation}
    \label{eq:T-gen}
    T^{MN}=(\rho_E+p_\vartheta)U^M U^N + (p_u - p_\vartheta) X^M X^N + (p_\chi-p_\vartheta) Y^M Y^N + p_\vartheta\, g^{MN} + \del^M{}_i\, \del^N{}_j\, \tau^{ij}\,, 
\end{equation}
where $U^M$ is the fluid's five-velocity, $X^M$ is a spacelike unit vector in the $u$-direction, $Y^M$ is a spacelike unit vector in the $\chi$-direction, while $p_u=p_u(u,\chi)$, $p_\chi=p_\chi(u,\chi)$, and $p_\vartheta=p_\vartheta(u,\chi)=p_\varphi$ are pressure components in the directions $u$, $\chi$, and $\vartheta$, respectively. 
Additionally, $\tau^{ij}$ is a traceless symmetric matrix describing the off-diagonal shear stress components of the stress-energy tensor.
For the quantities mentioned, it holds that
\begin{alignat}{2}
    &U^M = (U^t,0,0,0,0),  &&U^M U^N g_{MN}=-1\,,\\[2mm]
    &X^M = (0,X^u,0,0,0),  &&X^M X^N g_{MN}=1\,,\\[2mm]
    &Y^M = (0,0,Y^\chi,0,0),  &&Y^M Y^N g_{MN}=1\,,\\[2mm]
    &  \tau^{ij}=0,\, \forall\, (i,j)\neq \{(1,2), (2,1)\}, \hspace{1.5em} && \tau^{12}=\tau^{21}=\frac{3\rho'(u)}{\rho(u)}\left[ (\partial_\chi A) \left(1+\rho(u)\partial_\rho A \right)-\rho(u) \partial_\rho\partial_\chi A\right]\,.
\end{alignat}

It is interesting to note that since the warp function is of the form $A=A(\rho(u) \cos\chi)$, the differential equation $\tau^{12}=0$ can be solved explicitly.
By doing so, one realizes that shear stress vanishes altogether in the bulk, only in the case of the thin RS-II braneworld model, with warp function being given by $A(\rho,\chi)=-\ln\left(1+k \rho(u) |\cos\chi| \right)$.
Moreover, in the same model, one can readily show that the tangential pressures become all equal to each other, that is $p_\chi=p_\vartheta=p_\varphi$.
As a result, in this specific scenario, the stress-energy tensor reduces to an anisotropic fluid described by the following expression
\begin{equation}
    \label{eq:T-RSII}
    T^{MN}_{\rm RS-II}=(\rho_E+p_\vartheta)U^M U^N + (p_u - p_\vartheta) X^M X^N + p_\vartheta\, g^{MN}\,.
\end{equation}
For such a fluid, the energy conditions manifest themselves in the following way:
\begin{itemize}
    \item Null Energy Conditions (NEC): 
    \begin{equation}
        \label{eq:NEC}
        \rho_E+p_u\geq 0 \hspace{1em}  \text{\&} \hspace{1em} \rho_E+p_\vartheta\geq 0\,.
    \end{equation}
    \item Weak Energy Conditions (WEC):
    \begin{equation}
        \label{eq:WEC}
        \rho_E\geq 0 \hspace{1em}  \text{\&} \hspace{1em} \text{NEC}.
    \end{equation}
    \item Dominant Energy Conditions (DEC):
    \begin{equation}
        \label{eq:DEC}
        \rho_E-|p_u|\geq 0 \hspace{1em}  \text{\&} \hspace{1em} \rho_E - |p_\vartheta|\geq 0\,.  
    \end{equation}
    \item Strong Energy Conditions (SEC):
    \begin{equation}
        \label{eq:SEC}
        \text{NEC} \hspace{1em}  \text{\&} \hspace{1em} \rho_E+\frac{1}{2}(p_u+3p_\vartheta)\geq 0 \,.
    \end{equation}
\end{itemize}

For any other thick braneworld scenario, the stress-energy tensor will be given by \eqref{eq:T-gen} and, besides its additional anisotropy, it will always contain shear stress.
Consequently, the energy conditions are not as compactly expressed as before.
To obtain expressions similar to those derived earlier, one has to first diagonalize the stress-energy tensor.
This can be achieved either via an appropriate coordinate transformation or by introducing an orthogonal basis in a local Lorentz frame at each point in spacetime, following the procedure described in \cite{Maeda:2018hqu}. 
However, such an analysis lies beyond the scope of this article.
Our primary concern in this article and the subsequent examples is to present the geometrical features and characteristics of braneworld-uplifted wormholes in the bulk.
For each paradigm, we will analyze the energy conditions specifically within the RS-II model.
We anticipate similar conclusions to hold even in thick braneworld scenarios,
however, one should carefully examine each scenario individually to ensure robust and accurate results.

\section{Application to specific examples}
\label{Sec:Ex}

By utilizing the general methods developed in the previous section, we here provide specific examples for the uplift of various types of MT and nMT wormholes in both thin (RS-II) and thick braneworld models and subsequently study their higher-dimensional structures. In particular, we consider four well-known wormholes from the literature, covering cases with invertible and non-invertible metric functions $r(u)$, as well as with symmetry and asymmetry w.r.t.~the throat. The examples we consider are, on the one hand, the WHs of Casadio-Fabbri-Mazzacurati, and Bronnikov-Kim which are solutions to the effective 4D braneworld field equations and as such, provide typical examples in the context of the braneworld scenario. On the other hand, we consider wormholes that are not a priori associated with braneworld models, those are the Simpson-Visser spacetime, and the Ellis-Bronnikov or~\enquote{anti-Fisher} solution. In all of the aforementioned wormholes, we use the thin-brane RS-II braneworld model \cite{Randall:1999vf} as the main, common example for the uplift. Furthermore, in order to illustrate the effect of the braneworld model on the uplifted wormhole structure, using the Bronnikov-Kim wormhole as our brane seed metric, we provide a comparative study between the RS-II and a thick-brane model.

\subsection{Casadio-Fabbri-Mazzacurati braneworld wormholes}
\label{Sec:CFM}

Casadio, Fabbri and Mazzacurati (CFM), in their search for new black-hole solutions in the braneworld, obtained an interesting solution to the effective 4D brane equations that exhibits both black-hole and wormhole branches~\cite{Casadio:2001jg}. This metric had also been obtained by Germani and Maartens as a possible description of the spacetime outside a homogeneous star on the brane~\cite{Germani:2001du}.
In terms of the curvature radial coordinate the CFM brane metric is given by
\beq
\dd s_4^2=-\left(1-\frac{2M}{r}\right)\dd t^2+\frac{1-\frac{3M}{2r}}{\left(1-\frac{2M}{r}\right)\left(1-\frac{\alp}{r}\right)}\dd r^2+r^2\dd \Omega_2^2\,,
\label{eq:CFM_metric_4D}
\eeq
where $M\geqslant0$ is identified with the Arnowitt–Deser–Misner (ADM) mass \cite{Arnowitt:1961zz} of the solution. 
The value of the dimensionless ratio of parameters $\alp/(2M)>0$, determines the causal structure of~\eqref{eq:CFM_metric_4D} as follows:
\begin{itemize}
\item{$\alp/(2M) \in (1,+\infty)$ : Isometric traversable WH. Here $\alp$ is identified with the radius of the wormhole throat.}
\item{$\alp/(2M) = 1$ : WH/BH threshold.}
\item{$\alp/(2M) = 3/4$ : Schwarzschild BH.}
\item{$\alp/(2M) \in (0,3/4)$ : Schwarzschild-like BH with a spacelike curvature singularity located at $r=3M/2$.}
\end{itemize}
Here we will focus our analysis on the wormhole branch of the CFM spacetime.
However, the GEA will provide us with the consisted extension of~\eqref{eq:CFM_metric_4D} for any value of $\alp/(2M)$, and so, one may investigate the bulk properties of this interesting spacetime for any braneworld model.
In the representation \eqref{eq:CFM_metric_4D}, the coordinate location of the wormhole throat, is given by $r_{\rm th}=\alp$~\cite{Bronnikov:2021liv} and is identified with the physical radius of the throat since $r$ is the curvature radial coordinate.

Following the analysis of the preceding section, it is straightforward to deduce that the embedding of the CFM wormhole in the RS-II model, where $A(y)=-k|y|$ or $A(y(\rho \cos\chi))=-\ln(1+k\rho|\cos\chi|)$, is described by the following line element
\beq
\dd s^2=\frac{1}{(1+k\rho|\cos\chi|)^2}\Bigg[-\left(1-\frac{2M}{\rho}\right)\dd t^2+\frac{1-\frac{3M}{2\rho}}{\left(1-\frac{2M}{\rho}\right)\left(1-\frac{\alp}{\rho}\right)}\dd \rho^2+\rho^2\dd \Omega_3^2 \Bigg]\,,
\label{eq:CFM_metric_GEA}
\eeq
with $k>0$ corresponding to the inverse curvature radius in the bulk. As we have already discussed, after the embedding, the domain of the radial coordinate $r$ of the 4D wormhole, is inherited by the 5D radial coordinate $\rho$.
Therefore, the above line element describes the RS-II braneworld version of the CFM wormhole, iff $\alpha/(2M)>1$ and $\rho\in[\alpha,+\infty)$.
The domains for both the 4D radial coordinate $r$ and the bulk coordinate $z$ or $y$, are determined via the relation $\rho^2=r^2+z^2\geq \alpha^2$, with $r\in [0,+\infty)$ and $z\in (-\infty,+\infty)$.
The location of the hyperthroat stemming from the condition~\eqref{eq:Throat_location_GEA_MT} is given by $\rho=\alpha$, and thus the physical radius of the hyperthroat in the bulk, in accordance with Eq.~\eqref{eq:R_MT_GEA_generic}, is given by
\beq
R_{\rm eff}\left(\alp,\chi\right)=\frac{\alp}{1+k\alp|\cos\chi|}\,.
\label{eq:ReffCFM}
\eeq
The hyperthroat has a finite extend along the extra dimension, and its value can be readily obtained as
\beq
R_{\rm eff}\left(\alp,0\right)=R_{\rm eff}\left(\alp,\pi\right)=\frac{\alp}{1+k\alp}\,.
\eeq
In the~\enquote{weak bulk warping} limit $(k\alp\ll1)$, the $\chi$-dependence in~\eqref{eq:ReffCFM} is negligible, and so the 5D wormhole structure approaches spherical symmetry.

Out of the three parameters of~\eqref{eq:CFM_metric_GEA} $\{M,\alp, k\}$, we can construct various dimensionless combinations in order to study the properties of the spacetime. The combination $k\alpha$ in particular, is the most interesting one, since it includes the effect of the bulk curvature radius (via $k^{-1}$) and the wormhole throat curvature radius (via $\alpha$). As such, its value allows us to deduce the effect of the warping of the extra dimension on the wormhole structure for any fixed $\alpha$. The smaller the value of $k\alp$, the weaker the warping in the bulk (larger AdS$_5$ curvature radius). In our study of the 5D structure of spacetime we will thus use this dimensionless parameter for our analysis.

\begin{figure*}[t!]
\begin{center}
\includegraphics[width=0.49\linewidth]{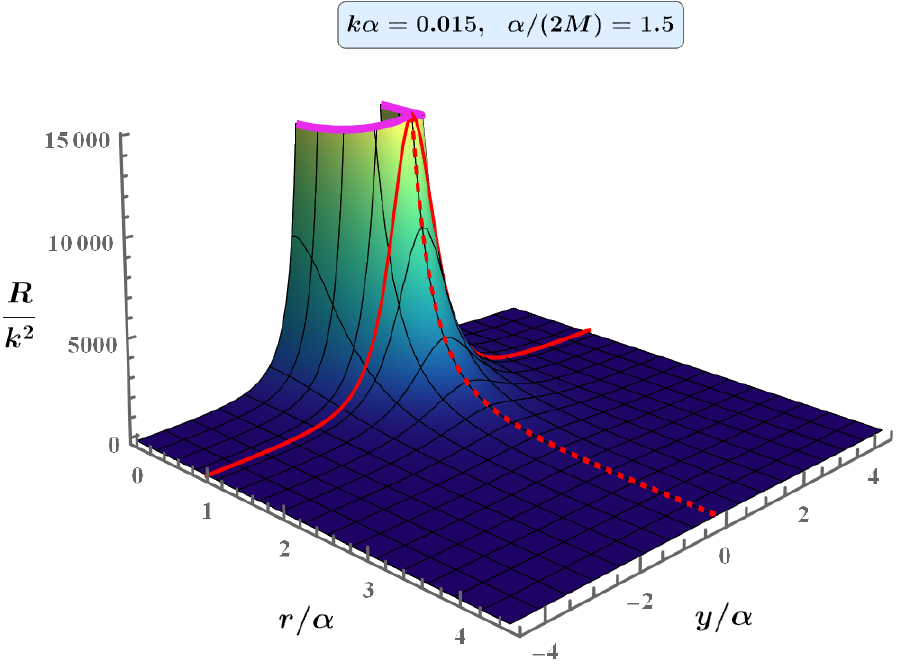}
\includegraphics[width=0.49\linewidth]{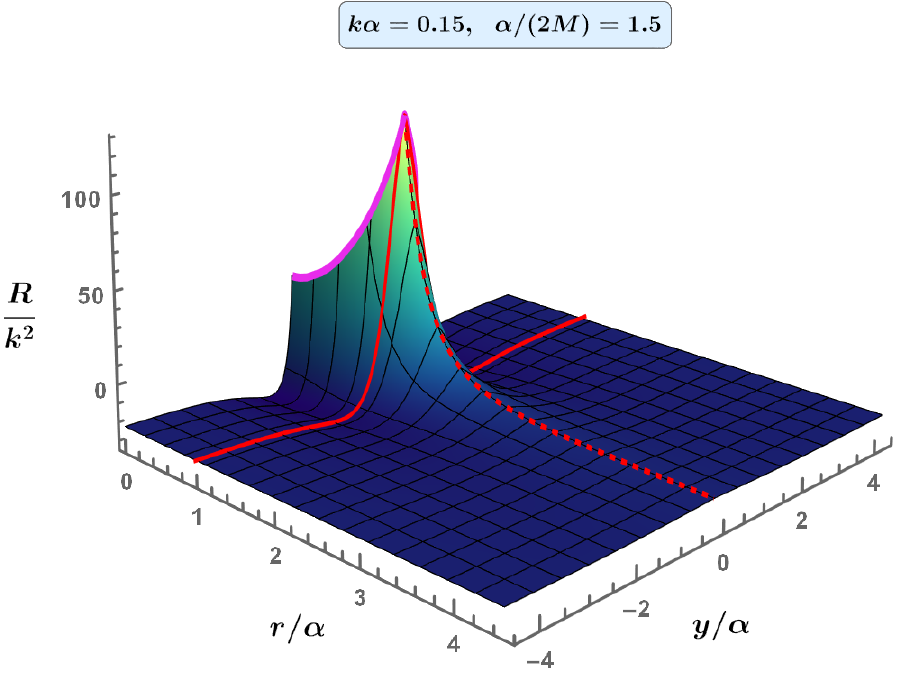}
\includegraphics[width=0.49\linewidth]{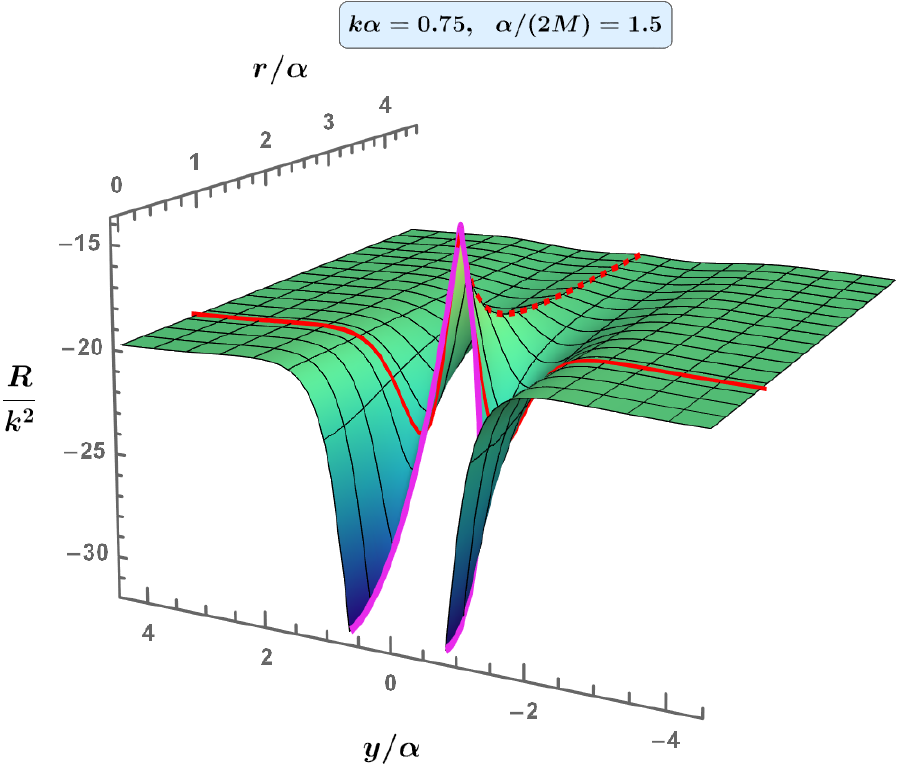}
\caption{The Ricci curvature of the Casadio-Fabbri-Mazzacurati wormhole embedded in the RS-II model for a fixed value of the dimensionless parameter $\alpha/(2M)=1.5$ and $k\alpha=\{0.015$, $0.15$, $0.75\}$. 
In all cases, the purple lines indicate the location of the wormhole throat as it extends into the bulk ($\rho=\alpha$), the continuous red line is for $r/\alpha=1$, while the dashed red line is for $y/\alpha=0$ (on the 3-brane).
All depicted quantities are dimensionless.}
\label{fig: CFM-R}
\end{center}
\end{figure*}

In Fig.~\ref{fig: CFM-R}, we depict the Ricci curvature $R$ for the line element~\eqref{eq:CFM_metric_GEA} for three different values of $k\alp$ and fixed value $\alpha/(2M)>1$.
%
%
We observe, that in all cases $R$ asymptotes to its pure AdS$_5$ value $R=-20 k^2$ very rapidly along the extra dimension, while it deviates from it only on and near the brane due to the curvature of the brane.
In all panels of Fig.~\ref{fig: CFM-R}, the purple lines indicate the location of the hyperthroat ($\rho=\alpha$) as it extends into the bulk.
Independently of the model parameters, the maximum value of the Ricci curvature takes place in the intersection between the solid red ($r/\alpha=1$) and dashed red lines ($y/\alpha=0$).
At this point, we meet the induced four-dimensional wormhole throat, $r=\alpha$, on the 3-brane. 
For any other point on the wormhole hyperthroat, we observe that $r<\alpha$.
This should not be surprising as it is a consequence of the braneworld uplift of the 4D wormhole geometry and Eq.~\eqref{eq:r_min_domain}.
This effect will become more clear during the upcoming discussion of the embedding diagrams for the current wormhole geometry.
In addition to the above, one notices that when the warping of the bulk is weak (top left panel), the spacetime curvature along the hyperthroat tends to be more homogeneous relative to its maximum value at the location of the wormhole throat on the brane.
This is similar to the case of a WH in flat-extra dimensions where the wormhole structure is globally spherically symmetric and as such the curvature on the hyperthroat is uniform along all dimensions. On the other hand, for mild bulk warping (top right panel) deviations of the curvature along the hyperthroat from its 4D limit become more prominent, while for strong bulk warping (bottom panel) the effect of the bulk has very significant impact on the curvature of the WH along the extra dimension. We also observe that the scalar curvature along the hyperthroat of the WH is not sign-definite, and it can be positive everywhere, negative everywhere or exhibiting regions of both, depending on the parameters. Similar findings regarding the sign of $R$ on the throat have also been reported for 4D WHs, see e.g.~\cite{Bakopoulos:2021liw}.

Pertaining now to the flare-out condition \eqref{eq:flare_out_r}, in the case of the braneworld CFM wormhole solution \eqref{eq:CFM_metric_GEA}, we obtain
\begin{equation}
\frac{e^{-A(y(z_0))}}{r}\left(\frac{2M}{\alpha}-1\right)\left(\frac{3M}{\alpha}-2\right)^{-1}>0\,,
\label{eq:CFM_GEA_flare_out_cond}
\end{equation}
for any braneworld model for which $e^{-2A(y)}>0,\ \forall\, y$ on the wormhole hyperthroat. Since the wormhole branch of the CFM spacetime is defined by $\alp/(2M)>1$, the above condition is always satisfied and as such we conclude that the bulk embedding of the four-dimensional CFM wormhole via the GEA preserves its wormhole structure for any braneworld model respecting the aforementioned conditions.

\begin{figure*}[t!]
\begin{center}
\includegraphics[width=0.49\linewidth]{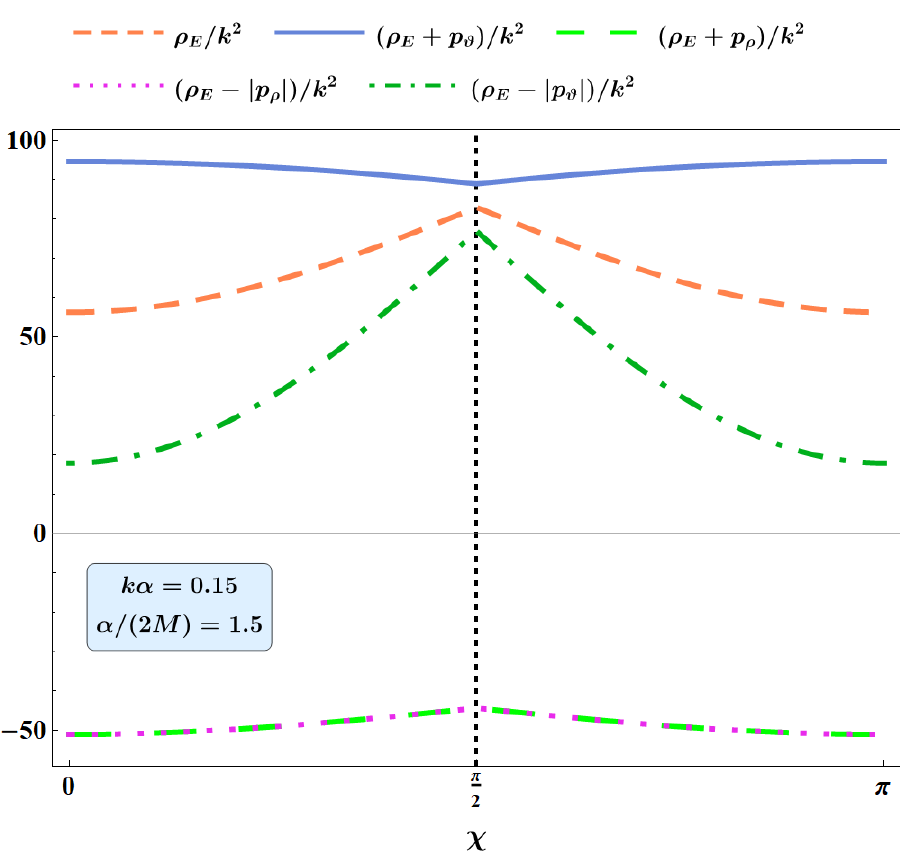}
\includegraphics[width=0.49\linewidth]{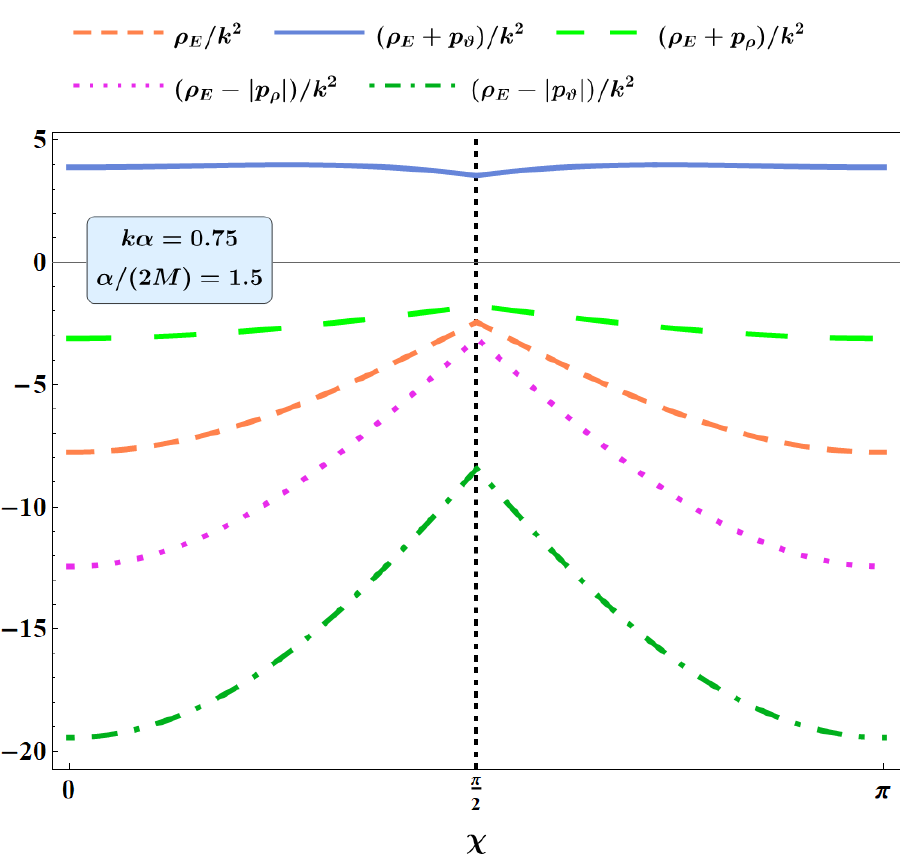}
\caption{The graphs of the quantities defining the energy conditions on the hyperthroat ($\rho=\alpha$) of the CFM wormhole embedded in the RS-II braneworld, in terms of the $\chi$-coordinate, for $\alpha/(2M)=1.5$ and $k\alpha=0.15$ (left panel), $0.75$ (right panel). All depicted quantities are dimensionless.}
\label{fig: CFM-en-con}
\end{center}
\end{figure*}

Returning now to the wormhole solutions embedded in the RS-II model, it becomes apparent that the energy conditions will be violated away from the brane due to the AdS$_5$ asymptotics.
Consequently, the most important region of spacetime to check the violation of the energy conditions is precisely along the wormhole hyperthroat that exhibits a finite extend into the bulk.
For the spacetime configuration described by \eqref{eq:CFM_metric_GEA} one finds that under the assumption \eqref{eq: act} the stress-energy tensor $T^{MN}$ is given by \eqref{eq:T-RSII} with the only difference being that now $X^M$ is a spacelike unit vector in the $\rho$-direction.
On the wormhole throat, where $\rho=\alpha$, the energy density $\rho_E$, the radial pressure $p_\rho$, and the tangential pressure $p_\vartheta$ are given respectively by
\begin{align}
    &\frac{\rho_E}{k^2}=-\frac{T^{t}{}_t}{k^2}=3\frac{(k\alpha) \left(3-5 \frac{M}{\alpha}\right) | \cos \chi| +2(k\alpha)^2 \left(2-3 \frac{M}{\alpha}\right)+\frac{M}{\alpha}-1}{(k\alpha)^2 \left(3 \frac{M}{\alpha}-2\right)}\,,
    \label{eq: CFM-rhoE}\\[2mm]
    &\frac{p_\rho}{k^2}=\frac{T^{\rho}{}_\rho}{k^2}= 6 + \frac{3 | \cos \chi| }{(k\alp)}-\frac{3}{(k\alp)^2}\,,
    \label{eq: CFM-pr}\\[2mm]
    &\frac{p_\vartheta}{k^2}=\frac{T^{\chi}{}_\chi}{k^2}=\frac{T^{\vartheta}{}_\vartheta}{k^2}=\frac{T^{\varphi}{}_\varphi}{k^2}=\frac{3}{2}\frac{ 2\left(3-4 \frac{M}{\alpha}\right) | \cos \chi| +(k\alpha) \left[\left(\frac{2M}{\alpha}-1\right) \cos (2\chi)-10\frac{M}{\alpha}+7\right]}{(k\alpha) \left(2-3\frac{M}{\alpha}\right)}\,.
    \label{eq: CFM-ptheta}
\end{align}
We have scaled all quantities in~Eqs.~\eqref{eq: CFM-rhoE}-\eqref{eq: CFM-ptheta} with $k^2$ to make them dimensionless. We see that the only independent variable, is the bulk coordinate $\chi\in[0,\pi]$ which encodes the deviations from the brane limit ($\chi=\pi/2$). On the brane, $\rho$ reduces to the four-dimensional radial coordinate $r$, and as a result, the throat radius becomes $r=\alpha$.

For the above anisotropic fluid, the energy conditions are given by Eqs.~\eqref{eq:NEC}-\eqref{eq:SEC} with $p_u$ replaced by $p_\rho$.
In Fig.~\ref{fig: CFM-en-con}, the energy conditions are depicted for the dimensionless parameters $\alp/(2M)=1.5$ and $k\alp=0.15,~0.75$. 
We observe that for mild bulk curvature ($k\alpha=0.15$), it is possible to satisfy only part of the energy conditions, while as the bulk curvature becomes more prominent ($k\alpha=0.75$), most of the conditions \eqref{eq:NEC}-\eqref{eq:SEC} are violated.
Specifically, for the quantity $\rho_E+p_\rho$ it holds that
\begin{equation}
    \frac{\rho_E+p_\rho}{k^2}=3 \left(1-\frac{2M}{\alpha}\right)\left( 3\frac{M}   {\alpha}-2\right)^{-1} \frac{1+(k\alpha) | \cos \chi|}{(k\alpha)^2}<0\,,
\end{equation}
which is always negative for RS-II braneworld CFM wormholes, since $\alpha/(2M)>1$ and $k\alpha>0$.
Consequently, the above analysis showcases that a theory of the form \eqref{eq: act} will always lead to the violation of null energy conditions.
To potentially avoid this violation, one would need to consider scalar-tensor theories with non-trivial couplings with gravity, and/or higher-order curvature or derivative terms.
Of course, such a task is highly non-trivial and goes beyond the scope of this article.

\begin{figure*}[t!]
\begin{center}
\includegraphics[width=0.49\linewidth]{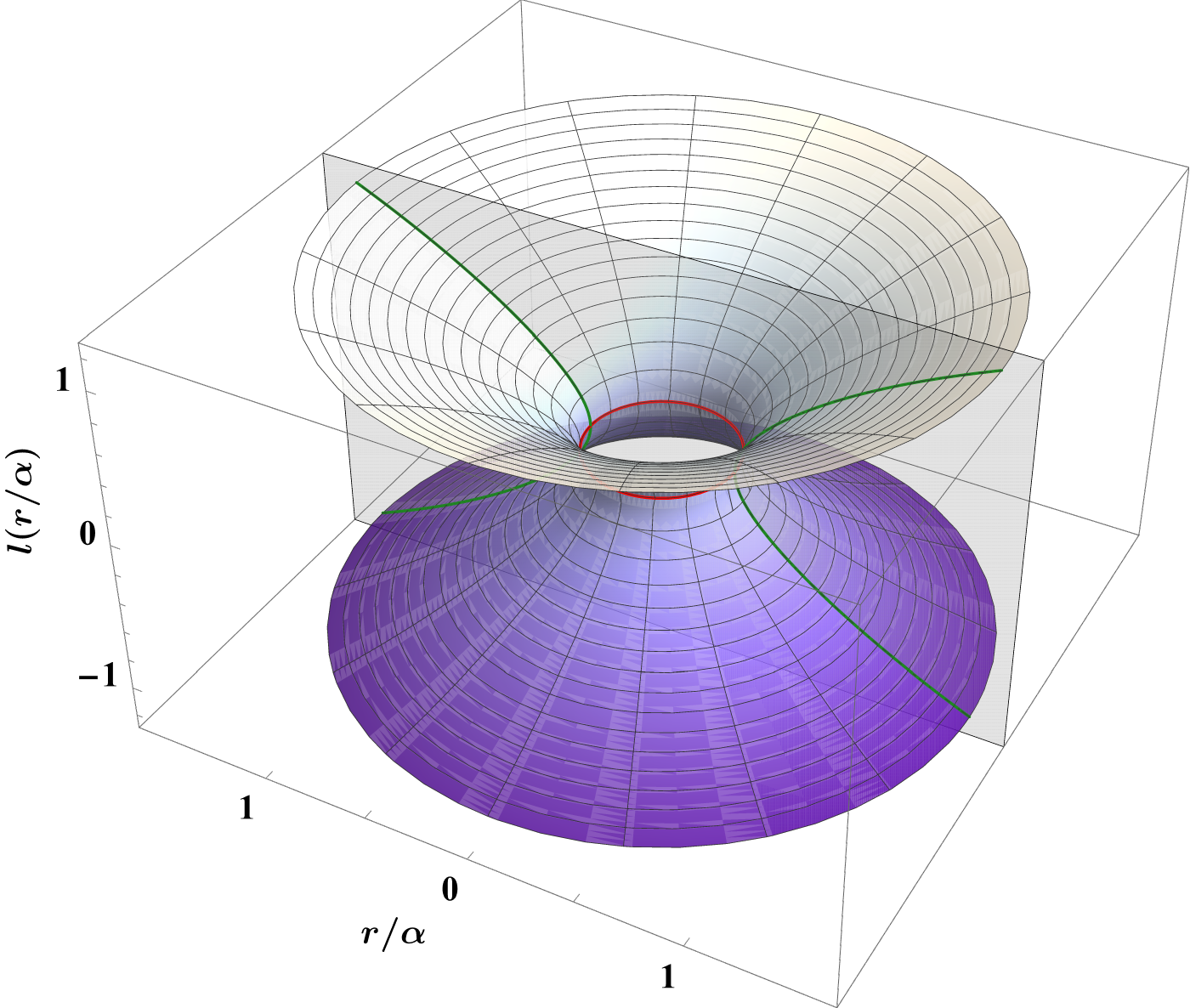}
\includegraphics[width=0.49\linewidth]{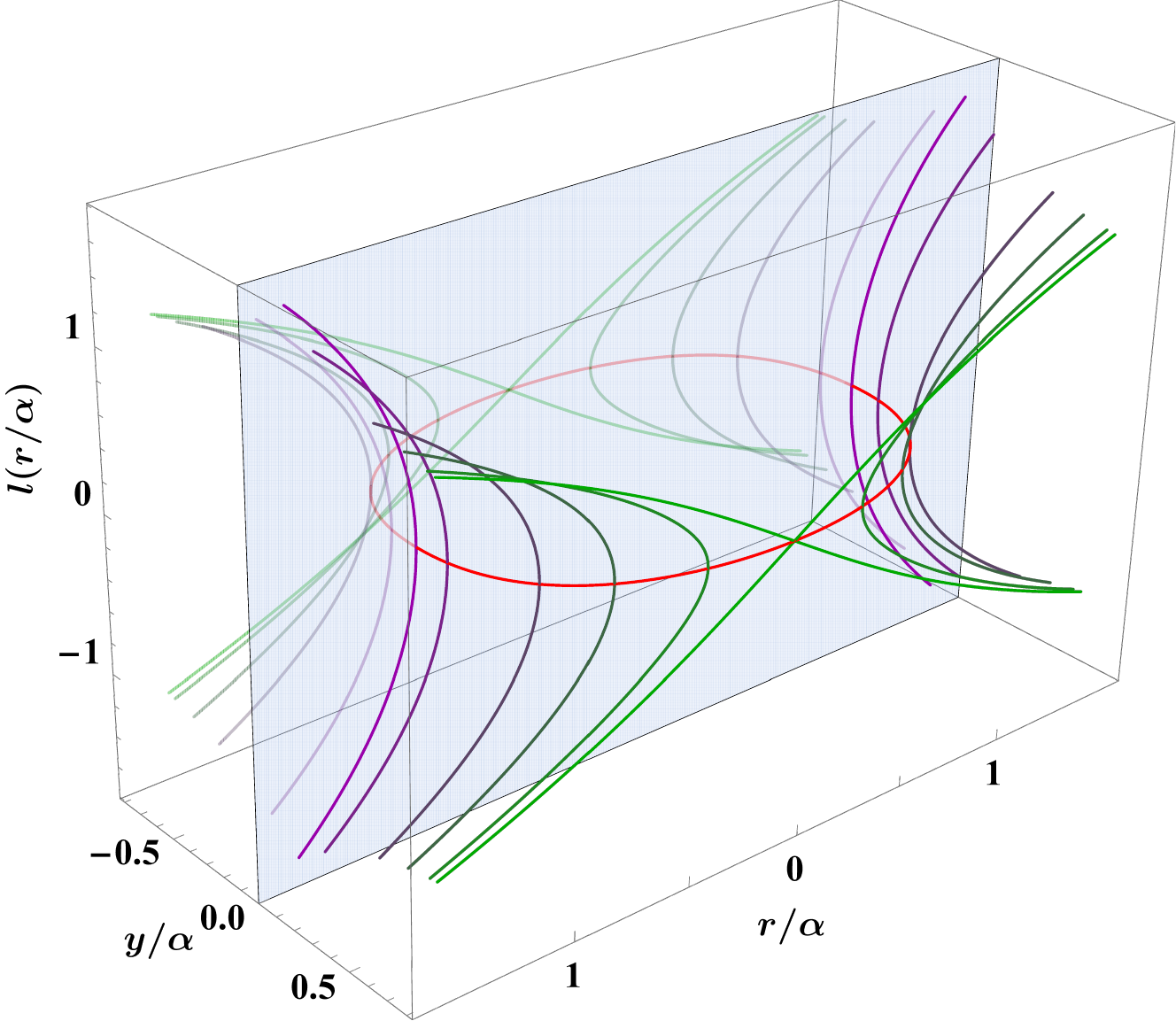}
\caption{For $k \alpha = 0.75$ and $\alp/(2M)=1.5$ the wormhole throat extends into the bulk up to $|y_{\rm max}|\simeq0.746\,\alpha$. Left panel: The embedding diagram for the bulk slice with $y_0=0.95\,y_{\rm max}$. The WH is spherically symmetric w.r.t. $r$ at each $y_0=\text{const.}$ and as such, the cross-sectional embedding diagram (CED) representation (green curves) encodes the complete information. Right panel: Embedding diagram representation of the 5D wormhole throat via stacks of CEDs at fixed $|y_0|$ spanning the circumference of the hyperthroat (red curve). The color coding of the curves is identical to those in~Fig.~\ref{fig:CFM_CEDs}. The light-blue plane represents the brane and all points interior to the throat circumference are not part of spacetime.}
\label{fig:3D_embedding_CFMRSII}
\end{center}
\end{figure*}

Let us finally discuss the embedding diagram representation of~\eqref{eq:CFM_metric_GEA}. In order to be able to illustrate 5D wormhole spacetimes via embedding diagrams, we need to utilize the symmetries of the spacetime to reduce the number of dimensions for the graphical representation. As we have discussed in Sec.~\ref{Sec:3}, each constant $y_0$ hypersurface in the bulk is spherically symmetric w.r.t. $r$.~This allows us to readily obtain the cross-sectional embedding diagram (CED) for the 5D WH at each $y_0=\text{const}.$ as for any static and spherically symmetric 4D wormhole spacetime by means of the lift function~\eqref{eq:lift_func_r} evaluated at $y_0$. Indeed, the CED can be seen in the left panel of Fig.~\ref{fig:3D_embedding_CFMRSII} and the hyperthroat projection on that hypersurface (red circle) is spherically symmetric as expected. 
At each $y_0$, the projected radius $r_{\rm th}$ of the wormhole throat is subject to the constraint~\eqref{eq:r_min_domain}. 
As a result, $r_{\rm th}< \alpha$ on each $y_0\neq 0$ hypersurface in the bulk, while $r_{\rm th}=\alpha$ on the brane ($y_0=0$).
Hence, the wormhole throat is shrinking as we move away from the brane. 
Due to the spherical symmetry of the $y_0$ bulk slicings, it is sufficient to only consider the 2D representation of the CED diagrams (green curves on left panel of Fig.~\ref{fig:3D_embedding_CFMRSII}), without loss of information about the embedding representation.

\begin{figure*}[t!]
\includegraphics[width=0.49\linewidth]{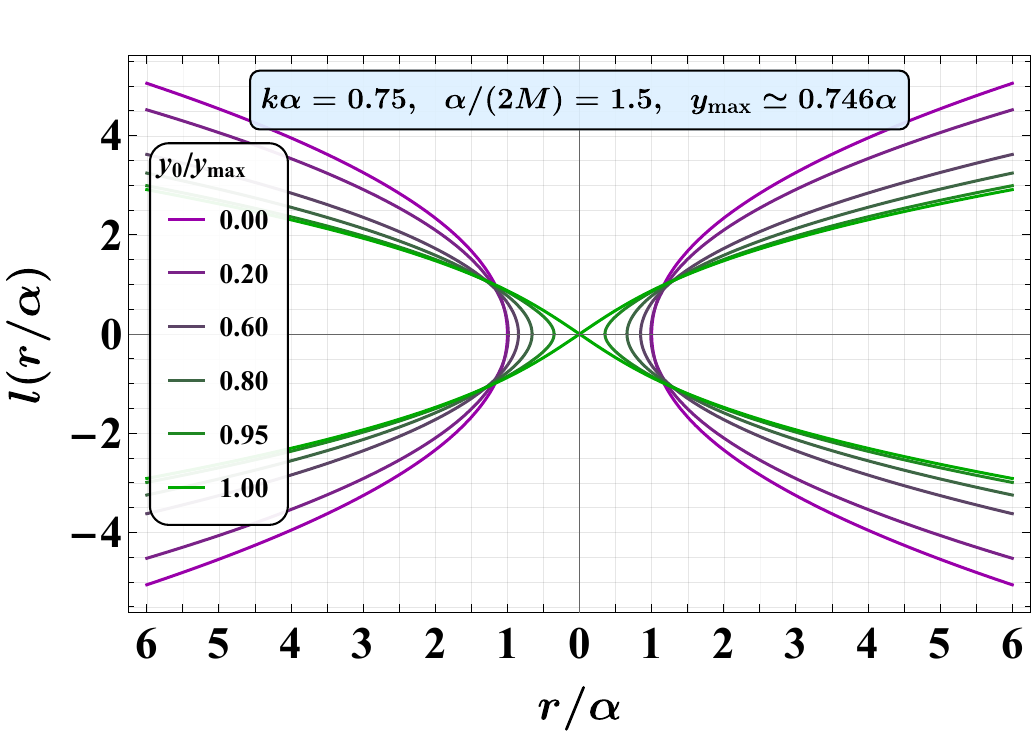}
\includegraphics[width=0.49\linewidth]{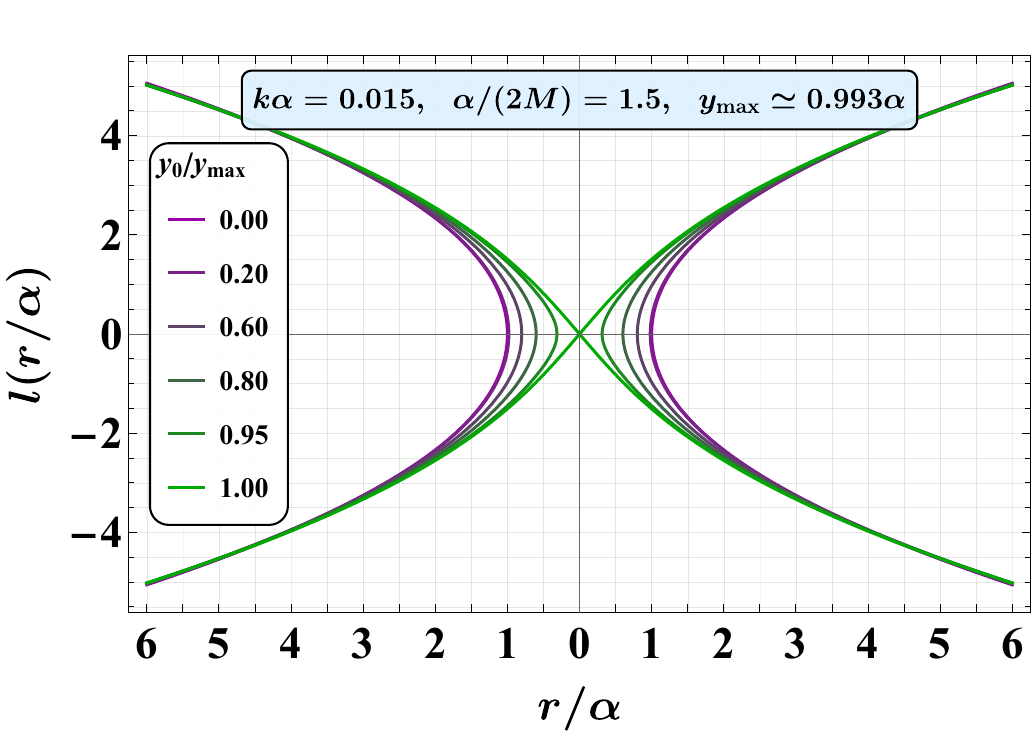}
\caption{Cross-sectional embedding diagrams (see Fig.~\ref{fig:3D_embedding_CFMRSII}) for the CFM WH in the RS-II braneworld model. The brane is located at $y_{\rm 0}=0$, and the hyperthroat extends to $y_{\rm 0}=\pm ymax$ in the bulk. The effect of the bulk curvature, parametrized here with the dimensionless variable $k \alp$, is to deform the WH along the extra dimension (left panel). For milder bulk curvature the WH is characterized by a more symmetric structure (right panel).}
\label{fig:CFM_CEDs}
\end{figure*}

Consequently, by considering stacks of constant $y_0$ slicings of the bulk we may use the 2D CEDs to obtain a representation of the complete 5D bulk wormhole structure as given in the right panel of Fig.~\ref{fig:3D_embedding_CFMRSII}. There, we can see that the 5D hyperthroat radius is not spherical due to the warping of the extra dimension. It shrinks monotonically as we move away from the brane, at an exponential rate, and vanishes at a distance from the brane equal to
\beq
|y_{\rm max}|=\left.\frac{\ln{\left(1+z_0 k\right)}}{k}\right\vert_{z_0=\alp}=\frac{\ln{\left(1+\alp k\right)}}{k}\,,
\label{eq:ymax_CFM_RSII}
\eeq
in the RS-II model that we consider here. For $y>y_{\rm max}$, constant $y_0$ hypersurfaces no longer 
exhibit wormhole topology since on their corresponding induced geometries there is no minimum surface area of constant radial coordinate $r$. However, the complete 5D spacetime preserves its wormhole structure, that remains localized near the brane. In the~\enquote{flat-extra dimension limit} of a vanishing bulk warping ($k \to 0$), we have from~\eqref{eq:ymax_CFM_RSII} that
\beq
|y_{\rm max}|=\alp-\frac{\alp^2 k}{2}+\mathcal{O}\left(k^2\right)\,,
\eeq
and thus we see that the hyperthroat in this limit approaches a spherically symmetric structure where $|y_{\rm max}|/a=1$.

On the left panel of Fig.~\ref{fig:CFM_CEDs}, we provide the exact same 5D embedding diagram as given on the right panel of Fig.~\ref{fig:3D_embedding_CFMRSII} in an equivalent 2D representation, with the extra dimension suppressed for a better quantitative understanding of the embedding diagrams. As the two panels of Fig.~\eqref{fig:CFM_CEDs} reveal, the effect of the bulk warping (parametrized once again in terms of the dimensionless parameter $k \alp$) is to deform the embedding diagram of the WH along the extra dimension. In the weak warping limit (right panel), the 5D structure of the embedding diagram is approaching a symmetric configuration reinforcing the observation of a~\enquote{flat extra dimension limit} in alignment with the analysis in terms of the curvature invariants above.

\subsection{Bronnikov-Kim braneworld wormholes}
\label{Sec:BK}

For our second example, we consider the so-called~\emph{zero-Schwarzschild mass} braneworld spacetime obtained by Bronnikov-Kim (BK)~\cite{Bronnikov:2002rn} as a solution to the effective four-dimensional braneworld field equations of Shiromizu-Maeda-Sasaki~\cite{Shiromizu:1999wj}, under the assumption of a vanishing 4-dimensional Ricci curvature. For this seed brane metric we will perform the uplift into two different braneworld models, one thin and one thick, and provide a comparative study between these two 5D geometries. The analysis in this section, highlights the differences and similarities of the 5D wormhole structures in different braneworld models.

The line element for the BK metric in the MT frame can be written as~\cite{Bronnikov:2021liv}
\beq
\dd s_4^2=-\left(1-\frac{\beta^2}{r^2} \right) \dd t^2+\left(1-\frac{\beta^2}{r^2} \right)^{-1}\left(1-\sqrt{\frac{2 \alp^2-\beta^2}{2 r^2-\beta^2}}\right)^{-1}\dd r^2+r^2 \dd \Omega_2^2\,.
\label{eq:BK_4D_metric}
\eeq
The two parameters $\alp$ and $\bet$ that appear in \eqref{eq:BK_4D_metric}, are integration constants with dimensions of $(\text{length})$. The former corresponds to the radius of the wormhole throat since at $r=\alp$ the $g_{11}$ metric component diverges, while the latter is constrained by $\alp>\beta$ in order for the WH to be traversable. Consequently, the domain of the radial coordinate is $r \in [\alp,+\infty)$. In the limiting case of $\alp/\beta \to 1$ the above geometry exhibits a BH/WH threshold with $\beta$ corresponding to the radius of the (double) BH event horizon.

The uplift of \eqref{eq:BK_4D_metric} by means of the GEA~\eqref{eq:GEA_rho_chi} leads to the line element
\beq
\dd s^2=e^{2A(y(\rho \cos\chi))}\Bigg[-\left(1-\frac{\beta^2}{\rho^2} \right) \dd t^2+\left(1-\frac{\beta^2}{\rho^2} \right)^{-1}\left(1-\sqrt{\frac{2 \alp^2-\beta^2}{2 \rho^2-\beta^2}}\right)^{-1}\dd \rho^2+\rho^2 \dd \Omega_3^2 \Bigg]\,,
\label{eq:BK_GEA_metric}
\eeq
with the warp function given by
\beq
A(y)=-k|y|\quad\Rightarrow\quad A(y(\rho \cos\chi))=-\ln(1+k\rho|\cos\chi|)\,,
\eeq
for the RS-II model, and by
\beq
\label{eq:thick-model}
A(y)=-\ln(\cosh(\sigma y))\quad\Rightarrow\quad A(y(\rho \cos\chi))=-\frac{1}{2}\ln(1+\sigma^2 \rho^2 \cos^2\chi)\,,
\eeq
for a thick brane model that is commonly considered in the literature (e.g.~\cite{Sui:2020atb,Moreira:2021cta,Tan:2022vfe, Li:2022kly, Moreira:2022zmx, Almeida:2023kfl}). The parameter $\sigma$, with units of $(\text{length})^{-1}$, corresponds to the curvature radius of AdS$_5$, and determines the thickness of the brane. As we have already discussed, after the uplift, the domain of the radial coordinate $r$ of the 4D wormhole, is inherited by the 5D radial coordinate $\rho$, hence, $\rho\in[\alpha,+\infty)$.

Since the BK WH is symmetric w.r.t. throat, the physical radius of the hyperthroat is determined solely by the warp factor~\eqref{eq:R_MT_GEA_generic} and as such, for the RS-II model, it will once again be given by~\eqref{eq:ReffCFM}, while for the thick-braneworld model under consideration, we have
\begin{equation}
R_{\rm eff}(\alpha,\chi)=\frac{\alp}{\sqrt{1+\sigma^2\alp^2\cos^2\chi}}\,.
\label{eq:Reff_thick}
\end{equation}
In the weak-warping limit, corresponding to $\sigma \alpha \ll 1$, the physical radius of the thick-brane model~\eqref{eq:Reff_thick} approaches once again a spherically-symmetric configuration in direct analogy with the thin-brane case.

\begin{figure*}[t!]
\includegraphics[width=0.49\linewidth]{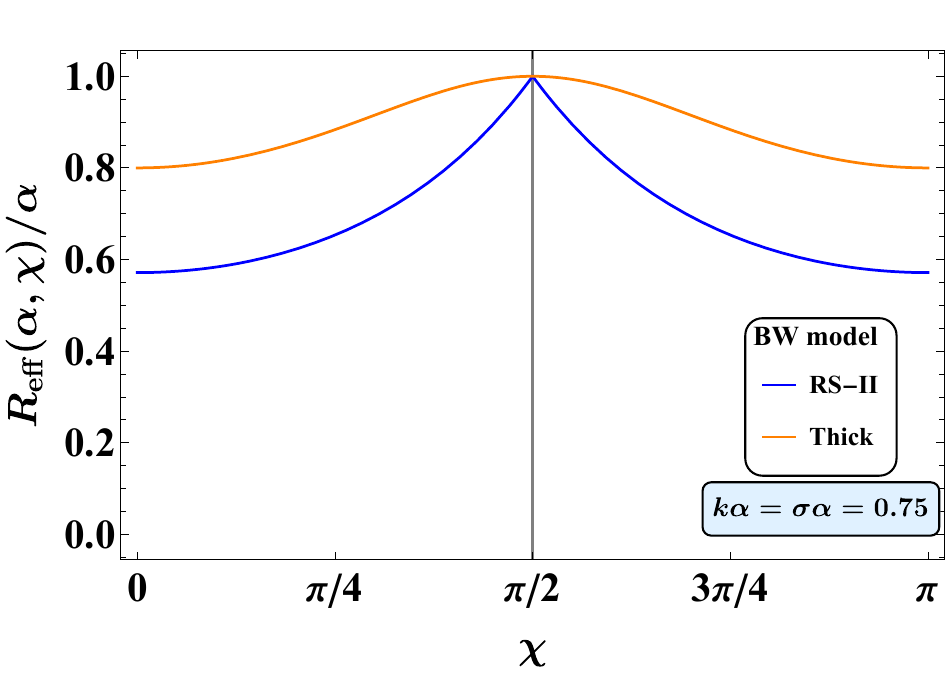}
\includegraphics[width=0.49\linewidth]{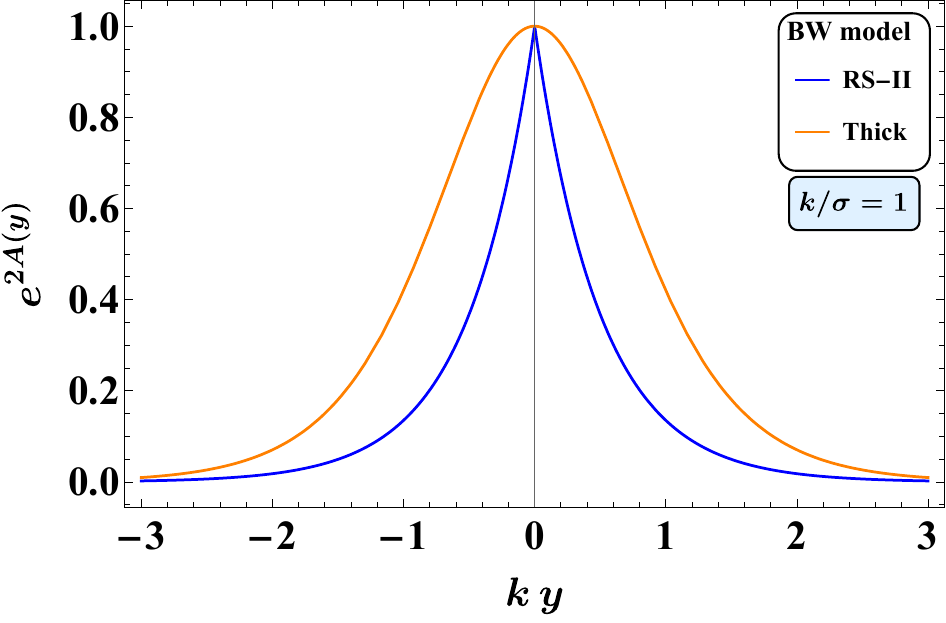}
\caption{The physical radius of the wormhole throat in the thick and thin braneworld models (left panel). The warp factors for the two braneworld models in the case of the same AdS$_5$ curvature radius in the bulk (right panel). All depicted quantities are dimensionless.}
\label{fig:Reff_thin_vs_thick}
\end{figure*}

The $\chi$-profile of the physical radius of the wormhole hyperthroat is shown on the left panel of Fig.~\ref{fig:Reff_thin_vs_thick} for the thick and thin braneworld models, in the case where the AdS$_5$ curvature radius in the bulk is the same in both models i.e. $k=\sigma$. A generic feature of thick-brane models is that they exhibit a milder warping near the brane in comparison with thin-brane models, as can be seen on the right panel of Fig.~\ref{fig:Reff_thin_vs_thick}. As a result, the WH structure in general deviates less from spherical symmetry in the vicinity of the brane for thick-brane models and this is also reflected in the profile of $R_{\rm eff}$. Independently of the braneworld model, the physical radius is maximized on the brane, and shrinks monotonically as we move away from the brane.

\begin{figure*}[t!]
\begin{center}
\includegraphics[width=0.49\linewidth]{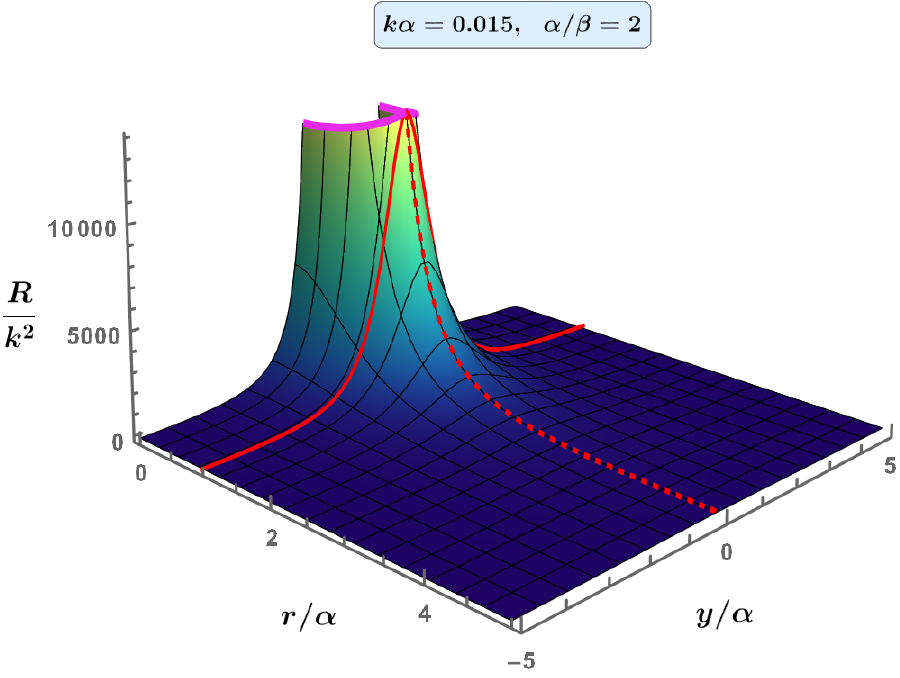}
\includegraphics[width=0.49\linewidth]{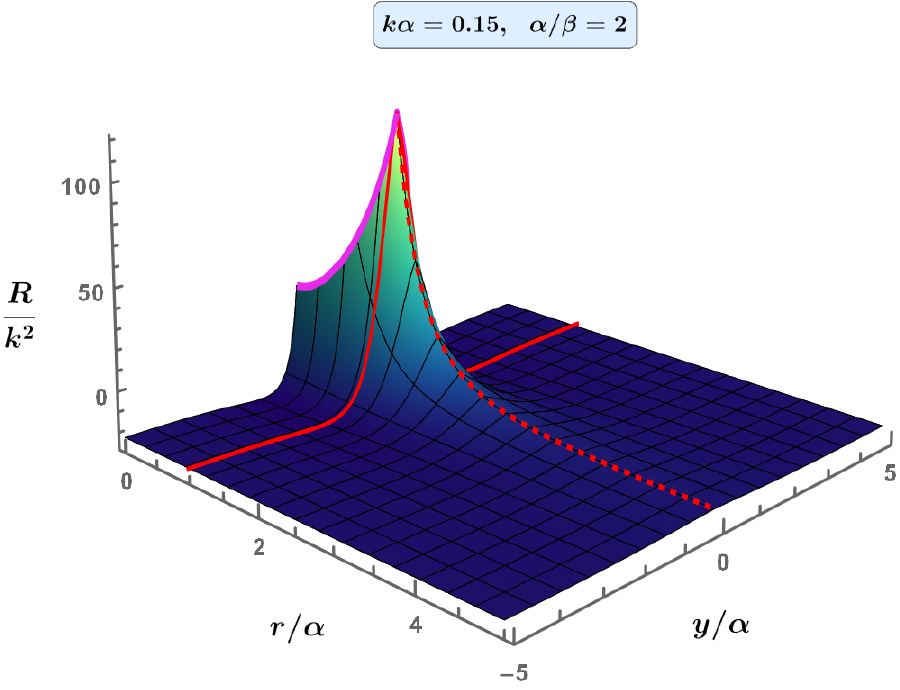}
\includegraphics[width=0.49\linewidth]{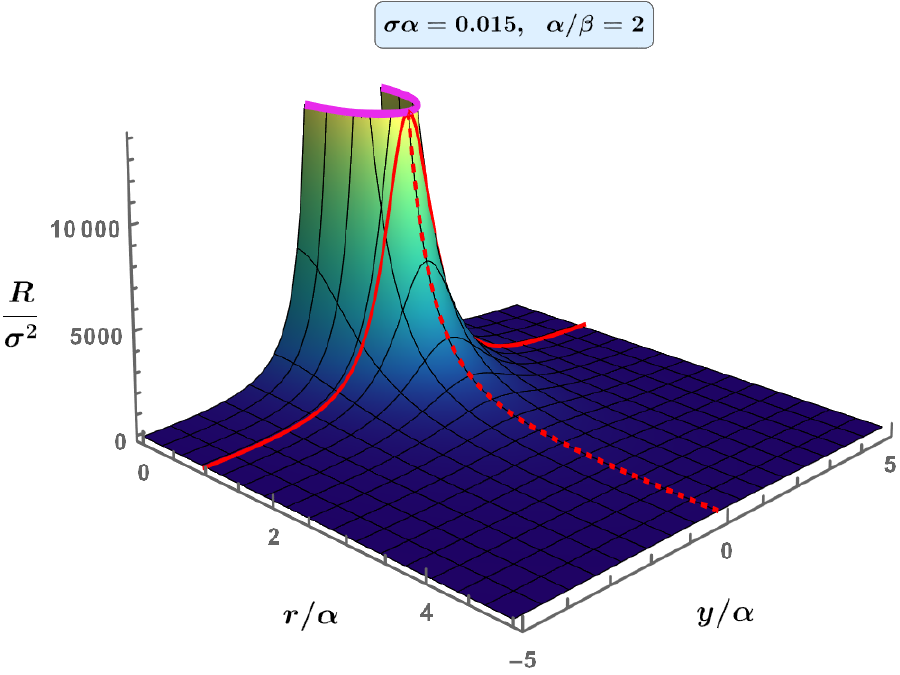}
\includegraphics[width=0.49\linewidth]{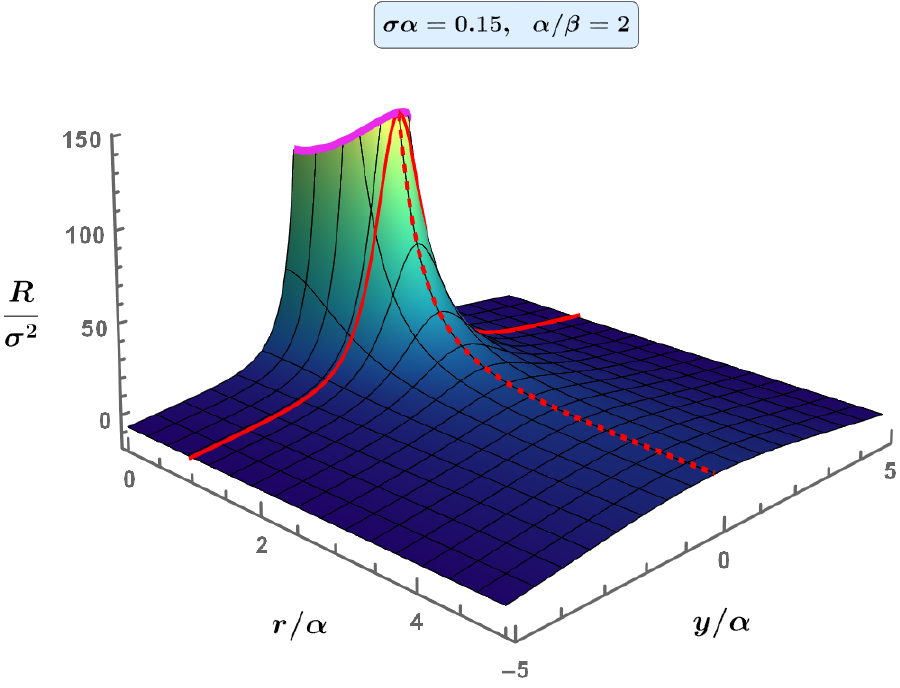}
\caption{The Ricci curvature of the Bronnikov-Kim wormhole embedded in the RS-II model (top panels) and the thick braneworld model \eqref{eq:thick-model} (bottom panels) for a fixed value of the dimensionless parameter $\alpha/\beta=1.5$ and varying $k\alpha$ and $\sigma\alpha$. 
In all cases, the purple lines indicate the location of the wormhole throat as it extends into the bulk ($\rho=\alpha$), the continuous red line is for $r/\alpha=1$, while the dashed red line is for $y/\alpha=0$ (on the 3-brane).
All depicted quantities are dimensionless.}
\label{fig: BK-R}
\end{center}
\end{figure*}

In the context of the RS-II model, by performing a similar analysis on the metric~\eqref{eq:BK_GEA_metric} as the one that was done for the braneworld Casadio-Fabbri-Mazzacurati wormhole \eqref{eq:CFM_metric_GEA}, one observes that the form of the Ricci curvature is nearly identical to that of the CFM case (compare Figs. \ref{fig: CFM-R} and \ref{fig: BK-R}).
This similarity extends to all curvature invariants which we omit here for brevity.
The qualitative behavior of the Ricci curvature for fixed $\alpha/\beta=2$ and $k\alpha=\{0.015,\, 0.15\}$, is the same as in Fig. \ref{fig: CFM-R} for the braneworld CFM wormhole.
However, for the thick braneworld model \eqref{eq:thick-model}, the Ricci curvature has a different profile in the bulk due to the difference in the warping of the extra dimension.
The main difference that one notices in Fig. \ref{fig: BK-R} is that for the thick braneworld model, the curvature at the vicinity of the brane is mainly affected by the seed wormhole geometry and not by the warping of the extra dimension. This is in alignment with the general property of the thick-brane models exhibiting a milder warping near the brane as discussed above.
Specifically, in the case of $\sigma\alpha=0.015$, where the warping of the extra dimension is very mild, the Ricci curvature at the wormhole throat remains nearly unchanged from brane to bulk. Finally, one may wonder about the fact that even though the seed brane-metric~\eqref{eq:BK_4D_metric} is an exact, Ricci-flat, solution to the effective Einstein equations on the brane, the uplifted five-dimensional geometry has a non-trivial Ricci curvature in the brane limit $\chi\to\pi/2$. Although this might seem inconsistent, it is in complete agreement with the analysis in \cite{Shiromizu:1999wj} and with Gauss' Theorema Egregium,
from which it can be shown that 
\begin{equation}
    \,^{(4)}R^{\mu}{}_{\nu\rho\sigma} \neq \,^{(5)}R^{\mu}{}_{\nu\rho\sigma}\big|_{\chi\rightarrow\pi/2}\,.
\end{equation}
In the above, $\,^{(4)}R^{\mu}{}_{\nu\rho\sigma}$ denotes the components of the Riemann tensor associated with the four-dimensional metric \eqref{eq:BK_4D_metric}, while $\,^{(5)}R^{\mu}{}_{\nu\rho\sigma}\big|_{\chi\rightarrow\pi/2}$ denotes the four-dimensional components of the five-dimensional Riemann tensor associated with the geometry \eqref{eq:BK_GEA_metric}, evaluated on the 3-brane.

For the metric~\eqref{eq:BK_GEA_metric}, the flare-out condition as evaluated by \eqref{eq:flare_out_r} leads to
\begin{equation}
\frac{e^{-A(y(z_0))}}{r}\left(1-\frac{\beta^2}{\alpha^2} \right)\left(2-\frac{\beta^2}{\alpha^2} \right)^{-1}>0\,.
\label{eq:BK_GEA_flare_out}
\end{equation}
Since the wormhole branch requires $\alp>\bet$, we observe that for any braneworld scenario that respects the constraint $e^{-A(y)}>0$, $\forall\, y$, Eq.~\eqref{eq:BK_GEA_flare_out} is satisfied everywhere for the BK WH.

Following the discussion in Sec.~\ref{Sec:En-conds}, we only consider here the energy conditions for the thin-brane model, which, similarly to the example of the previous section, are found to be violated in this case as well under the assumption~\eqref{eq: act}.
The expressions for the energy density, the radial and the tangential pressures are given below
\begin{align}
    &\frac{\rho_E}{k^2}= 3\,\frac{ \left(2 \beta^2/\alpha^2-3\right) (k\alpha) | \cos \chi | +2 \left(\beta^2/\alpha^2-2\right) (k\alpha)^2+1}{\left(2-\beta^2/\alpha^2\right) (k\alpha)^2}\,, \\[2mm]
    &\frac{p_\rho}{k^2}= 6 + \frac{3 | \cos \chi| }{(k\alp)}-\frac{3}{(k\alp)^2}\,, \\[2mm]
    &\frac{p_\vartheta}{k^2}=3\,\frac{2 \left(\beta^2/\alpha^2-3\right) | \cos \chi | +(k\alpha) \left[\left(1-\beta^2/\alpha^2\right) \cos (2 \chi )+3 \beta^2/\alpha^2-7\right]}{2 \left(\beta^2/\alpha^2-2\right) (k\alpha)}\,.
\end{align}
For the same values of the dimensionless parameter $k\alpha$ and $\alpha/\beta=1.5$, one can verify that the corresponding graphs determining the energy conditions for a braneworld BK wormhole are almost identical to the ones depicted in Fig.~\ref{fig: CFM-en-con}.
Moreover, it can be shown explicitly that
\begin{equation}
    \frac{\rho_E+p_\rho}{k^2}=-\frac{3 \left(\beta^2/\alpha^2-1\right) [1+(k\alpha) | \cos \chi|]}{\left(\beta^2/\alpha^2-2\right) (k\alpha)^2}<0\,.
\end{equation}

\begin{figure*}[t!]
\begin{center}
\includegraphics[width=0.49\linewidth]{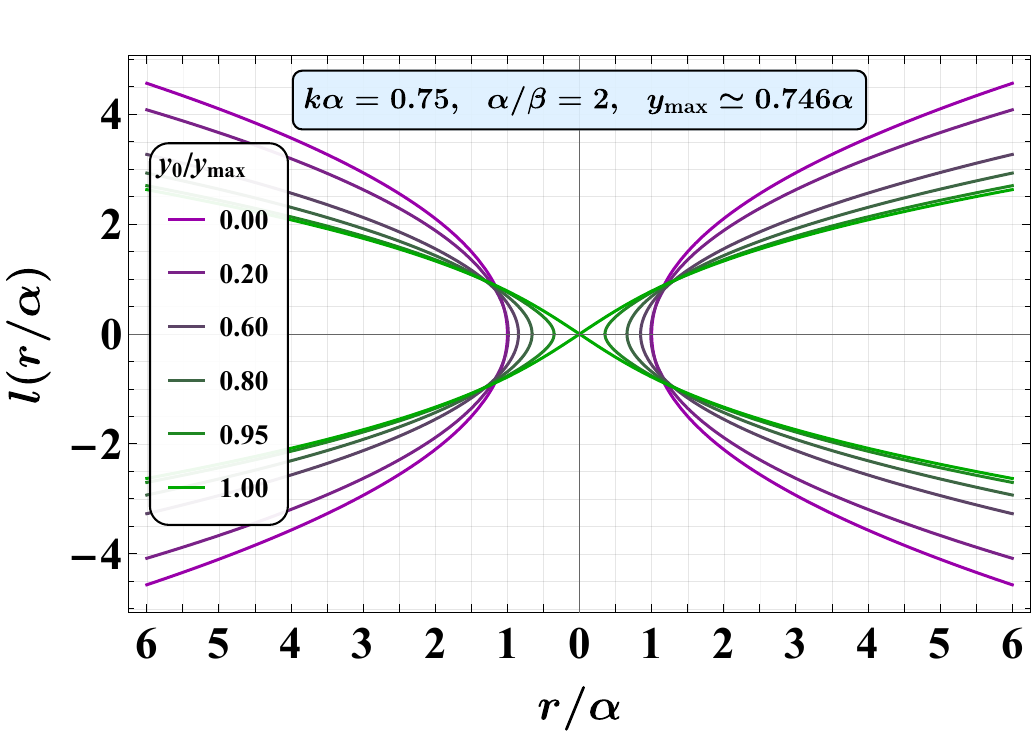}
\includegraphics[width=0.49\linewidth]{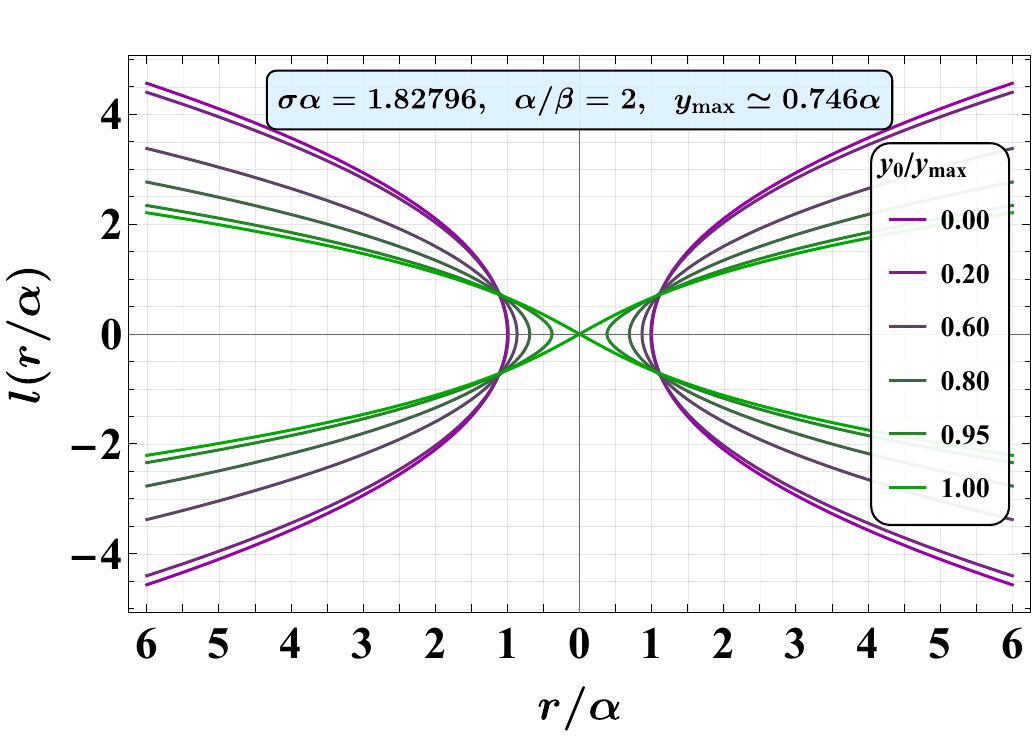}
\includegraphics[width=0.49\linewidth]{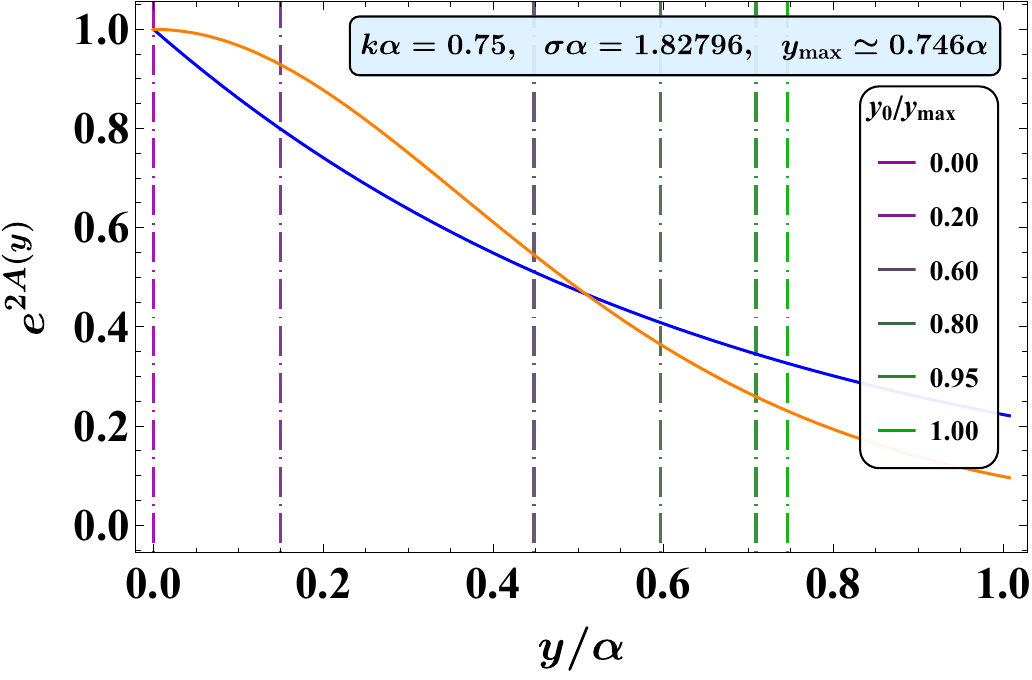}
\includegraphics[width=0.49\linewidth]{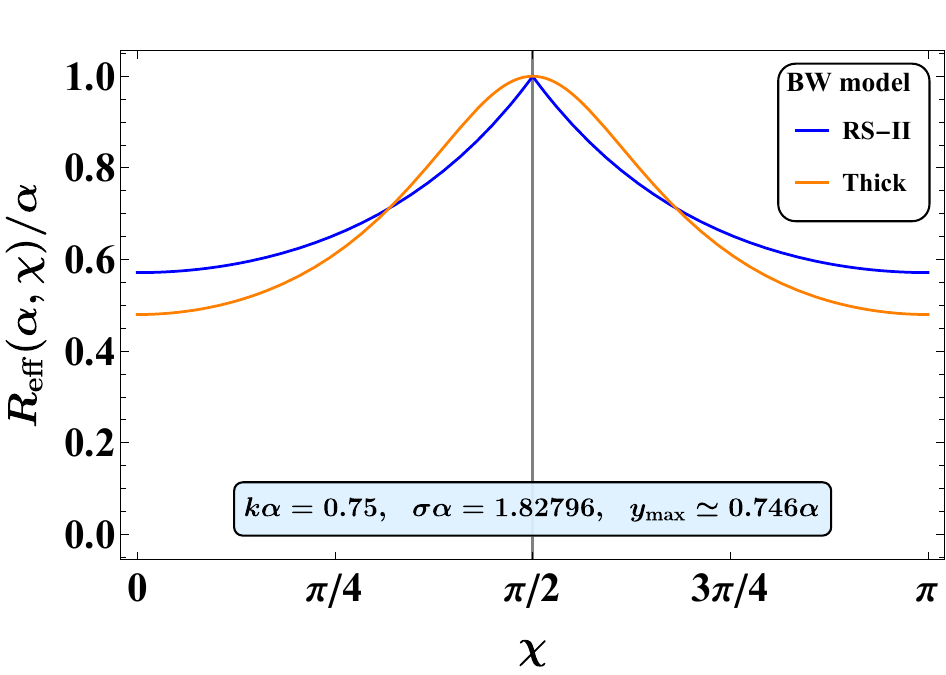}
\caption{Cross-sectional embedding diagrams (CEDs) (see Fig.~\ref{fig:3D_embedding_CFMRSII}) for the BK WH in the RS-II braneworld model (top left), and in the thick-brane model~\eqref{eq:thick-model} (top right). The warp factors (bottom left) for the RS-II (blue) and thick-brane (orange) model. The dashed lines represent the locations of the various bulk slicings considered for the CEDs in the top figures. The corresponding $\chi$-profile for the physical radius of the hyperthroat for throat-symmetric WHs in the RS-II model~\eqref{eq:ReffCFM} vs.~the thick brane model~\eqref{eq:Reff_thick} (bottom right).}
\label{fig:BK_CEDs}
\end{center}
\end{figure*}

Turning now to the study of the cross-sectional embedding diagrams (CEDs), in order to facilitate the comparison between the two braneworld models we choose values for the parameters $k$ and $\sigma$ such that the WH hyperthroat exhibits the same maximum value for the coordinate $y$ in both cases. For the RS-II model, $y_{max}$ is given by~\eqref{eq:ymax_CFM_RSII} while for the thick-brane model we have
\beq
|y_{\rm max}|=\left.\frac{\arcsinh{\left(z_0 \sigma\right)}}{\sigma}\right\vert_{z_0=\alp}=\frac{\arcsinh{\left(\alp \sigma\right)}}{\sigma}\,.
\label{eq:ymax_thick}
\eeq
Fixing the braneworld model parameters in this way, allows us to clearly identify the impact that the different warping of the extra dimension has on the CEDs as shown in Fig.~\ref{fig:BK_CEDs}. In the top two panels of the figure, we present the CEDs for various constant $y_0$ slicings in the bulk for the RS-II model on the left, and the thick-brane model~\eqref{eq:thick-model} on the right. On the bottom left panel we give the $y$-profiles of the warp factors for the two braneworld models and we denote the coordinate locations for the various bulk slicings that appear in the CED plots with the dashed colored lines, following the same color-coding with the top-panel figures. By comparing those three panels one can see that the profiles of the warp factors directly affect the amount of deviation of the bulk CEDs from the brane CED at $y_0=0$. In particular, for near-brane bulk slicings, where the warping is milder for the thick-brane model, the corresponding CEDs deviate less from the brane CED than in the thin-brane case. However, as we move further into the bulk, and for the particular choice of parameters used in this example, we see that the warping intensity of the thick-brane model becomes stronger than the thin-brane one for $y_0/y_{\rm max}\gtrapprox 0.509$ and this is also reflected in the CEDs. Finally, in the bottom right panel of Fig.~\ref{fig:BK_CEDs}, we see that for small deviations from the brane-limit~\footnote{Notice that in thick-brane models the brane is not located strictly at $\chi=\pi/2$ but extends along the extra dimensions without definite boundaries. With this observation in mind, in the context of the GEA the~\enquote{brane-limit} of $\chi \to \pi/2$ for thick branes can be thought of a referring to the~\enquote{center} of the brane defined as the spacelike hypersurface on which the warp function vanishes.} $\chi\to\pi/2$, the physical radius in the case of the thick-brane model exhibits a milder deviation from a spherically symmetric configuration, while moving deeper into the bulk where $\chi \to 0,\pi$, the physical radius is more intensely compressed along the extra dimension for the thick-brane model, relative to the thin-brane case, a behavior that is once again in alignment with the $y$-profiles of the warp factors.

\subsection{Simpson-Visser wormholes}
\label{Sec:SV}

In this subsection, we turn to the uplift of nMT seed brane metrics, and consider a simple toy-model with an invertible circumferential radius metric function $r(u)$, ensuring a straightforward transformation between frames. This allows us to provide a clear demonstration of the equivalence between the GEA and eGEA approaches for the uplift when $r(u)$ is invertible. On the other hand, for seed-metrics with non-invertible $r(u)$, such as in the case studied in the next subsection, this equivalence no longer holds, and the eGEA remains as the only known consistent method for the uplift of nMT seed metrics (see also Fig.~\ref{Fig: algorithm}).

A regularization method to remove the curvature singularities at the center of BHs has been proposed by Simpson and Visser (SV) in~\cite{Simpson:2018tsi}. The regularization is performed at the level of the metric and the field theory supporting the regularized geometry has to be obtained independently. When the SV method is applied to the Schwarzschild solution of mass $M$, the so-called SV metric is obtained in the nMT frame~\cite{Simpson:2018tsi}
\beq
\dd s_4^2=-\left(1-\frac{2M}{\sqrt{u^2+\alp^2}} \right)\dd t^2+\left(1-\frac{2M}{\sqrt{u^2+\alp^2}}\right)^{-1}\dd u^2+\left(u^2+\alp^2\right) \dd \Omega^2_2\,,
\label{SV_met}
\eeq
where $\alp \in [0,+\infty)$ is a regularization parameter with dimension $(\text{length})$, and the domain of the radial coordinate is $u \in (-\infty,+\infty)$. Depending on the value of $\alp$ relative to $M$, the SV metric can describe a BH ($\alp=0$), a black bounce ($\alp<2M$), a one-way WH with an extremal null throat ($\alp=2M$) or a two-way traversable Lorentzian WH ($\alp>2M$).

The spacetime~\eqref{SV_met} has attracted a lot of attention, with its properties and generalizations being investigated in a number of works~\cite{Mazza:2021rgq,Bronnikov:2021uta,Bronnikov:2021liv}. The SV regularization procedure has also been applied to a plethora of other metrics, see e.g.~\cite{Lima:2023jtl} and references therein. Inspired by the SV approach, an alternative regularization method a la Bardeen has been proposed in~\cite{Bronnikov:2024izh}. Recently, in an attempt towards a generalization of the SV metric to the Randall-Sundrum II (RS-II)~\cite{Randall:1999vf} braneworld model, the authors of~\cite{Crispim:2024yjz} applied the SV regularization method to the black-string solution of CHR~\cite{Chamblin:1999by}. Their findings reveal that even though this approach yields the SV geometry on the brane, it exhibits curvature singularities at the AdS horizon~\cite{Crispim:2024yjz}. 
The Nakas-Kanti approach for localized SV wormholes has been followed in \cite{Crispim:2024nou}, where the authors studied the case of a RS-II braneworld model.

In the wormhole branch of the line element~\eqref{SV_met}, the coordinate location of the throat is determined by the surface-area minimization condition
\beq
r'(u)= \frac{u}{\sqrt{u^2+\alp^2}}=0 \Rightarrow u_{\rm th}=0\,,
\eeq
along with the extremization condition ensuring the existence of a throat at $u_{\rm th}$
\beq
r''(u_{\rm th})=\alp^2>0\,.
\eeq
The physical radius in the nMT frame is then given by
\beq
R_{\Join}=r(u_{\rm th})=\alp\,.
\eeq
The MT-frame representation of~\eqref{SV_met} can be obtained under the radial coordinate redefinition $u \rightarrow r(u) : r^2 = u^2+\alp^2$, with $ r \in [\alp,+\infty)$
\beq
\dd s_4^2=-\left( 1-\frac{2M}{r}\right) \dd t^2+\left( 1-\frac{2M}{r}\right)^{-1} \left( 1-\frac{\alp^2}{r^2}\right)^{-1} \dd r^2+r^2 \dd \Omega^2_2\,.
\label{SV_met_MT}
\eeq
In this representation, the coordinate location of the wormhole throat is determined via~\eqref{eq:MT_u_th_general} to be $r_{\rm th}=\alp$, and is identified with the physical radius of the throat since $r$ is the curvature radial coordinate in this frame. 

Following the GEA, it is straightforward to obtain the 5D MT extension of \eqref{SV_met_MT} in the RS-II model, given by the line element
\beq
\label{eq: SV-RS2-metr}
\dd s^2=\frac{1}{(1+k\rho|\cos\chi|)^2}\Bigg[-\left( 1-\frac{2M}{\rho}\right) \dd t^2+\left( 1-\frac{2M}{\rho}\right)^{-1} \left( 1-\frac{\alp^2}{\rho^2}\right)^{-1} \dd \rho^2+\rho^2 \dd \Omega^2_3 \Bigg]\,,
\eeq
with $k>0$, $\rho\in[\alpha,+\infty)$, ($\alpha>0$). Since~\eqref{eq: SV-RS2-metr} is throat-symmetric, the physical radius of the hyperthroat is given once again by~\eqref{eq:ReffCFM}. Using \eqref{eq:flare_out_r}, one can readily find that the flare-out condition for \eqref{eq: SV-RS2-metr} is evaluated to be
\begin{equation}
\frac{e^{-A(y(z_0))}}{r}\left(1-\frac{M}{\alpha}\right)>0\,,
\label{eq:SV_GEA_flare_out_cond}
\end{equation}
for any braneworld model with $e^{-2A(y)}>0,\ \forall\, y$ thus ensuring that the uplift of the SV metric preserves its WH structure in the bulk for these models. Regarding the embedding diagrams for the RS-II braneworld SV wormhole, they are qualitatively very similar to in the ones already presented in Figs.~\ref{fig:CFM_CEDs}~and~\ref{fig:BK_CEDs}.

In the nMT frame representation, the application of the eGEA for the RS-II model with the seed brane metric~\eqref{SV_met} results via~\eqref{eq:ds2_eGEA} to the line-element
\beq
\\d s^2=\frac{1}{\left(1+k\sqrt{u^2+\alp^2}|\cos\chi|\right)^2}\Bigg[-\left( 1-\frac{2M}{\sqrt{u^2+\alp^2}}\right) \dd t^2+\left( 1-\frac{2M}{\sqrt{u^2+\alp^2}}\right)^{-1} \dd u^2+\left(u^2+\alp^2\right) \dd \Omega^2_3 \Bigg]\,.
\label{eq: SV-RS2-metr_eGEA}
\eeq
From the general extremization~\eqref{eq:nMT_extr_cond2} and minimization~\eqref{eq:nMT_min_con} conditions applied for the line element~\eqref{eq: SV-RS2-metr_eGEA}, we find that the coordinate location of the WH hyperthroat in the nMT frame is given by $u_{\rm th}=0$. Consequently, the physical radius is readily calculated to be~\eqref{eq:Reff-nMT}
\beq
R_{\rm eff}\left(u_{\rm th},\chi\right)=\left.\frac{\sqrt{u^2+\alp^2}}{\left(1+k\sqrt{u^2+\alp^2}|\cos{\chi}|\right)}\right\vert_{u=u_{\rm th}}=\frac{\alp}{1+k\alp |\cos{\chi}|}\,,
\eeq
which is identical to~\eqref{eq:ReffCFM} as it should, since~\eqref{eq: SV-RS2-metr} and~\eqref{eq: SV-RS2-metr_eGEA} are different representations of the same bulk spacetime.~It should also be noted that the five-dimensional representations of the uplifted metric~\eqref{eq: SV-RS2-metr} and~\eqref{eq: SV-RS2-metr_eGEA} are related via the radial coordinate transformation $u \to \rho(u)=\sqrt{u^2+\alp^2}$, that reduces to the relation between the four-dimensional radial coordinates of the two frames in the brane limit, see~\eqref{eq:r_rho_eGEA_bridge}.

The study of curvature invariants and the energy conditions for the uplifted RS-II SV wormhole can be performed in either of the two frames. Regarding the former,~it is easy to verify that for the same dimensionless parameters $k\alpha$ and $\alpha/(2M)$, the profile of the Ricci curvature obtained from \eqref{eq: SV-RS2-metr} has the same qualitative characteristics as in Fig.~\ref{fig: CFM-R}. As for the energy conditions, the relevant quantities on the wormhole hyperthroat are evaluated to be
\begin{align}
    &\frac{\rho_E}{k^2}=3\,\frac{(k\alp) \left(\frac{2M}{\alp}-2\right) |\cos\chi| -2 (k\alp)^2+\frac{2M}{\alp}}{(k\alp)^2}\,,\label{eq:rho_E_SV_GEA} \\[2mm]
    &\frac{p_\rho}{k^2}=6+ \frac{3 | \cos \chi| }{(k\alp)}-\frac{3}{(k\alp)^2}\,,\label{eq:p_r_SV_GEA} \\[2mm]
    &\frac{p_\vartheta}{k^2}=5+\frac{5 | \cos \chi | }{(k\alp)}+\frac{1}{(k\alp)^2}\left(1-3\frac{M}{\alp} \right)+\left(\frac{3}{2} \frac{M}{\alp}-1\right) \cos (2 \chi )+\frac{3}{2} \frac{M}{\alp}\,.\label{eq:p_theta_SV_GEA}
\end{align}
Notice that the sum of the energy density $\rho_E$ and the radial pressure $p_\rho$ is given by the following relation
\begin{equation}
    \frac{\rho_E+p_\rho}{k^2}=3\Bigg(\frac{2M}{\alpha}-1\Bigg)\frac{1+(k\alpha)|\cos\chi|}{(k\alpha)^2}\,,
\end{equation}
and is negative-definite on the hyperthroat for all values of $\chi\in[0,\pi]$.
Consequently, under the assumption \eqref{eq: act}, the energy conditions are violated on the hyperthroat.

Overall, our analysis reveals, that under the uplift to a given braneworld model, the Casadio-Fabbri-Mazzacurati, Bronnikov-Kim and Simpson-Visser wormholes exhibit qualitatively the same physical characteristics in the bulk, and this highlights the important role of the braneworld model for the uplift.

\subsection{Ellis-Bronnikov (anti-Fisher) wormholes}
\label{Sec:EB}

For our final example, we consider a seed brane metric that is non-isometric about the throat and is characterized by a non-invertible circumferential radius metric function, for which the uplift can only be performed by means of the eGEA. The analysis in this subsection illustrates the impact that the asymmetry of the seed brane metric has on the structure of the uplifted five-dimensional wormholes.

The field equations of General Relativity with a minimally-coupled and massless scalar field, admit as a solution a spacetime corresponding to a naked singularity, that was first derived by Fisher in 1948~\cite{Fisher:1948yn}. In 1973, Ellis and Bronnikov independently realized that if the kinetic term of the scalar field is taken to have the opposite sign (phantom field), the solution for the metric tensor corresponds to a traversable WH spacetime that became know as the Ellis-Bronnikov (EB) solution~\cite{Ellis:1973yv,Bronnikov:1973fh}. The generalization of this solution to include rotation and the study of its properties have been investigated numerically in~\cite{Chew:2016epf}. The EB WH is oftentimes also referred to in the literature as the~\enquote{anti-Fisher} solution, in analogy with the~\enquote{anti-de Sitter} solution. The line element for the EB WH can be written as~\cite{Kashargin:2007mm,Bronnikov:2021liv}
\beq
\dd s_4^2=-e^{2w(u)} \dd t^2+e^{-2w(u)}\left[\dd u^2+\left(u^2+\bet^2 \right) \dd \Omega^2_2 \right]\,,\qquad w(u)=\frac{\alp}{\bet}\left[\arctan\left(\frac{u}{\bet}\right)-\frac{\pi}{2} \right]\,,
\label{eq:ds2_Ellis-Bronnikov}
\eeq
where $\alp$ and $\bet$ are integration constants with dimensions of $(\text{length})$. The absence of event horizons in spacetime, so that the WH traversability requirement is fulfilled, demands that $w(u)$ is finite $\forall\, u \in (-\infty,+\infty)$ and this entails that $\beta \neq 0$. The later constraint on $\bet$ also ensures the absence of curvature singularities. For the metric~\eqref{eq:ds2_Ellis-Bronnikov}, the circumferential radius metric function is~\emph{non-invertible} and it is given by
\beq
r(u)=e^{-w(u)}\sqrt{u^2+\bet^2}\,.
\label{eq:r(u)_EB}
\eeq 
According to our discussion in Sec.~\ref{Sec:2}, the coordinate location of the throat is determined by the extremization condition~\eqref{eq:r0_cond_1} $r'(u_{\rm th})=0$, and corresponds to $u_{\rm th}=\alp$. The minimization condition~\eqref{eq:r0_cond_2} $r''(u_{\rm th})>0$ is satisfied for any finite value of $\alp$ and $\bet$ ensuring that the throat is always located at $u_{\rm th}=\alp$. The physical radius of the throat is then given by
\beq
R_{\Join}\equiv r(u_{\rm th})=e^{-w(\alp)}\sqrt{\alp^2+\bet^2}\,.
\eeq
It should be noted that, for any $\alp>0$ the two sides of the wormhole, corresponding to the intervals $u \in (-\infty,\alp)$ and $u \in [\alp,+\infty)$ of the radial coordinate domain, are non-isometric according to~\eqref{eq:ds2_Ellis-Bronnikov} and as such, the wormhole is said to be~\emph{throat asymmetric}. In the limiting case of $\alp=0$ the wormhole becomes throat-symmetric and the physical radius reduces to $R_{\Join}=\sqrt{\beta^2}>0$.

Since~\eqref{eq:r(u)_EB} is non-invertible, the uplift of~\eqref{eq:ds2_Ellis-Bronnikov} into the bulk is only possible in terms of the extended formulation of the GEA (Sec.~\ref{Sec:eGEA}), and thus, for the RS-II model the resultant 5D geometry obtained from~\eqref{eq:ds2_eGEA} is
\beq
\dd s^2=\frac{1}{\left(1+ke^{-w(u)}\sqrt{u^2+\bet^2}|\cos\chi|\right)^2}\Bigg[-e^{2w(u)} \dd t^2+e^{-2w(u)}\dd u^2+e^{-2w(u)}\left(u^2+\bet^2\right) \dd \Omega^2_3\Bigg]\,,
\label{eq:ds2_aF_eGEA}
\eeq
with
\beq
w(u)=\frac{\alp}{\bet}\left[\arctan\left(\frac{u}{\bet}\right)-\frac{\pi}{2} \right]\,.
\eeq
The extremization and minimization conditions~\eqref{eq:nMT_extr_cond2} and~\eqref{eq:nMT_min_con} when applied for~\eqref{eq:ds2_aF_eGEA} yield the coordinate location of the hyperthroat $u_{\rm th}=\alp$ and so its physical radius is evaluated via~\eqref{eq:Reff-nMT} to be
\beq
R_{\rm eff}\left(u_{\rm th},\chi\right)=\left.\frac{\sqrt{u^2+\bet^2}}{\left(e^{w(u)}+k\sqrt{u^2+\bet^2}|\cos{\chi}|\right)}\right\vert_{u=u_{\rm th}}=\frac{\sqrt{\alp^2+\bet^2}}{e^{w(\alp)}+k\sqrt{\alp^2+\bet^2}|\cos{\chi}|}\,.
\label{eq:R_eff_EB_asymm}
\eeq
Notice that, unsurprisingly,~\eqref{eq:R_eff_EB_asymm} shares the same functional form with~\eqref{eq:ReffCFM}, since the uplift is performed in the same braneworld model. However, due to the non-isometry of~\eqref{eq:ds2_Ellis-Bronnikov}, $R_{\rm eff}$ now also depends on the metric functions of the seed brane metric. This is in contrast to~\eqref{eq:ReffCFM} that remains the same for any isometric wormhole uplifted in the RS-II model. Indeed, one can see that in the isometric limit of $\alpha=0$,~\eqref{eq:R_eff_EB_asymm} also becomes independent of the details of the brane geometry and it is identified once again with~\eqref{eq:ReffCFM}.

\begin{figure*}[ht!]
\begin{center}
\includegraphics[width=0.49\linewidth]{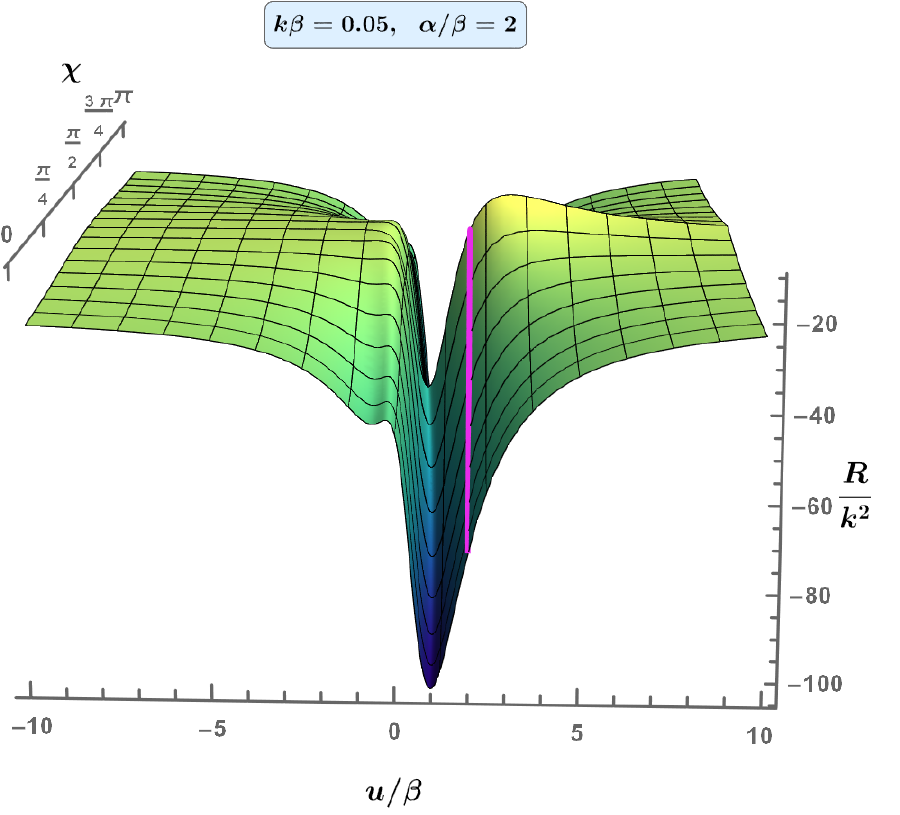}
\includegraphics[width=0.49\linewidth]{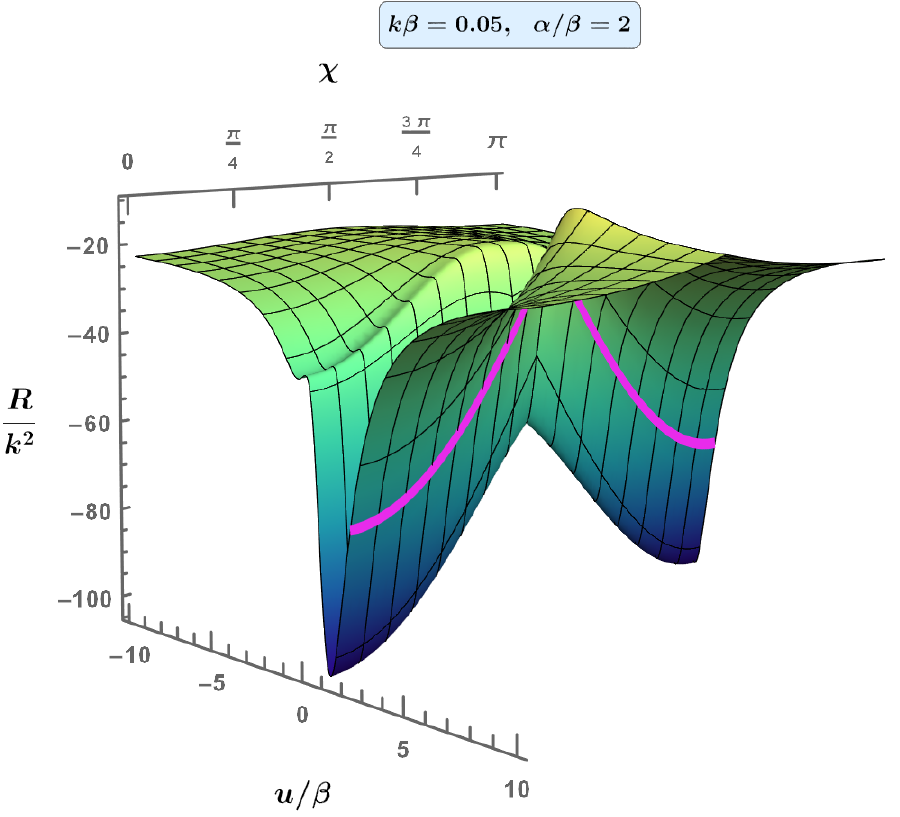}
\includegraphics[width=0.49\linewidth]{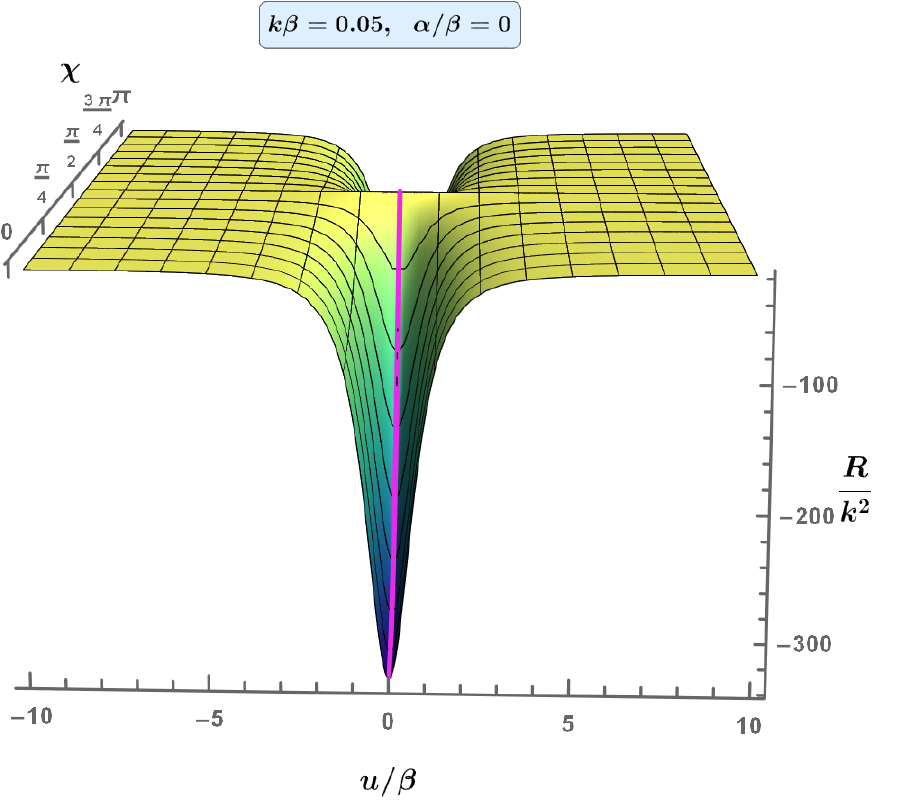}
\includegraphics[width=0.49\linewidth]{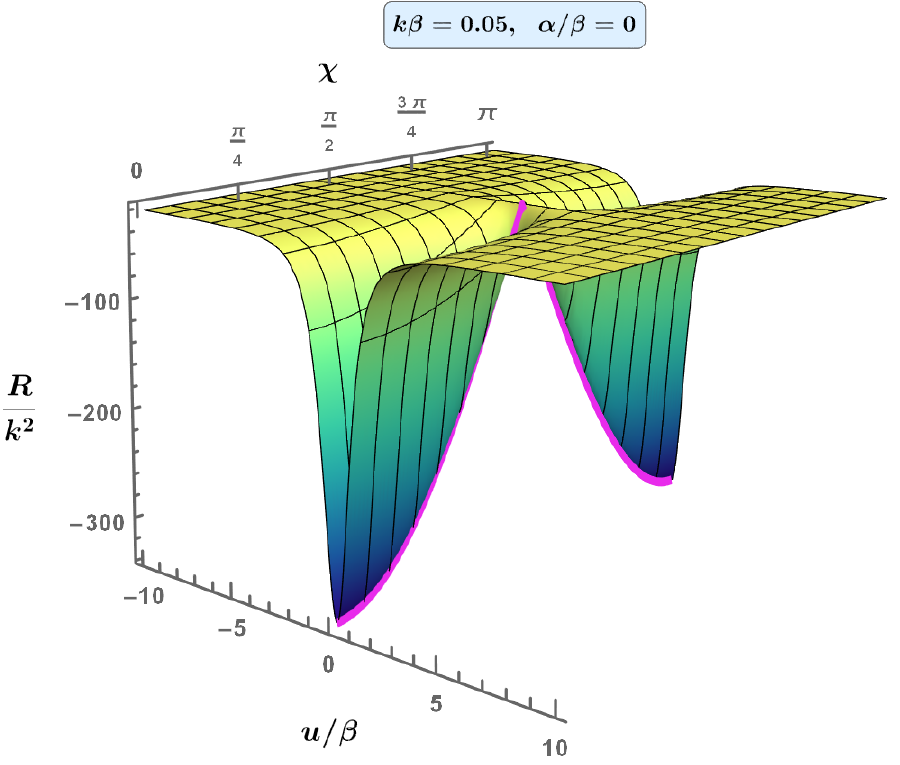}
\caption{The Ricci curvature of the Ellis-Bronnikov wormhole embedded in the RS-II model. Both asymmetric and symmetric hyperthroat cases are shown, top graphs and bottom graphs, respectively,
for the dimensionless parameters $\alpha/\beta=2$ and $k\beta=0.05$. 
In all cases, the purple lines indicate the location of the wormhole hyperthroat at $u=\alpha$.
All depicted quantities are dimensionless.}
\label{fig: EB-R}
\end{center}
\end{figure*}

As one of the general properties of the eGEA discussed in Sec.~\ref{Sec:eGEA}, the $\bold{Z}_2$ bulk symmetry of the RS-II model in~\eqref{eq:ds2_aF_eGEA} remains unaffected by the non-isometry of the seed-brane metric. This is also reflected in the $\chi$-profile of the physical radius~\eqref{eq:R_eff_EB_asymm} and the 5D Ricci curvature for~\eqref{eq:ds2_aF_eGEA} which are depicted in Fig.~\ref{fig: EB-R}. The first observation to be made is that in all cases, the Ricci scalar\footnote{And for that matter all other curvature invariants as well.} quickly approaches its asymptotic AdS$_5$ value in the bulk, when moving away from the brane, which, in Fig.~\ref{fig: EB-R} is along the direction of the edges of the plots where $\chi=0,\pi$, and $u\to\pm\infty$. For the non-isometric case ($\alp\neq0$), presented in the top panels of Fig.~\ref{fig: EB-R}, the Ricci curvature is reflection-asymmetric with respect to the hyperthroat, and furthermore, at each $\chi=\text{const.}$ spacelike hypersurface, the throat is not located at an extremum of the curvature. On the other hand, in the isometric limit ($\alp=0$), depicted in the bottom panels of Fig.~\ref{fig: EB-R}, the Ricci scalar becomes symmetric about the hyperthroat, and the latter is located at the minimum of the curvature for each $\chi=\text{const.}$ slicing of the bulk.

The aforementioned consequences of the non-isometry of the wormhole and its impact on the five-dimensional curvature invariants and the location of the hyperthroat are in direct analogy with the situation in 4D. We may thus conclude that any throat-asymmetry characterizing the seed-brane metric is also reflected in the structure of the uplifted bulk geometry. 
However, in properly interpreting the correlation between the seed-brane metric and its uplift, one must always have in mind the distinction between the 4D Ricci curvature on a $\chi_0=\text{const.}$ hypersurface and the 5D Ricci curvature evaluated at the same $\chi_0$.
The reason behind this has been discussed in Sec. \ref{Sec:BK}, while a thorough analysis regarding this matter can be found in \cite{Shiromizu:1999wj}.

Lastly, regarding the energy conditions in the case of the line-element~\eqref{eq:ds2_aF_eGEA}, one can readily determine that on the wormhole's hyperthroat the relevant quantities are given by
\begin{align}
    &\frac{\rho_E}{k^2}=-6\left(1+\frac{e^{w(\alpha)} | \cos \chi| }{k\sqrt{\alpha^2+\beta^2}}\right)\,,
    \label{eq:rho_E_EB_eGEA} \\[2mm]
    &\frac{p_\rho}{k^2}=6-\frac{3 e^{2 w(\alpha )}}{k^2 \left(\alpha ^2+\beta ^2\right)}+\frac{3 e^{w(\alpha )} | \cos \chi | }{k \sqrt{\alpha ^2+\beta ^2}}\,,\label{eq:p_r_EB_eGEA} \\[2mm]
    &\frac{p_\vartheta}{k^2}=5+\frac{e^{2 w(\alpha )}}{k^2 \left(\alpha ^2+\beta ^2\right)}+\frac{5 e^{w(\alpha )} | \cos \chi | }{k \sqrt{\alpha ^2+\beta ^2}}-\cos (2 \chi
   )\,.\label{eq:p_theta_EB_eGEA}
\end{align}
Using the above, it is straightforward for one to show that the energy conditions cannot be satisfied in the bulk. 
Specifically, close to the brane, where $\chi\rightarrow\pi/2$, we find that
\begin{equation}
    \frac{\rho_E+p_\rho}{k^2}\bigg|_{\chi\rightarrow \pi/2}=-\frac{3e^{2w(\alpha)}}{k^2(\alpha^2+\beta^2)}\,,
\end{equation}
therefore, the NEC are violated.

\section{Conclusions}
\label{Sec:Conclusions}

By utilizing the recently-proposed General Embedding Algorithm (GEA)~\cite{Nakas:2023yhj}, one can uplift any static spherically symmetric 4D geometry, into any 5D single-brane braneworld model characterized by a non-singular and injective warp factor. However, one must be careful when uplifting brane geometries with non-trivial topology such as in the case of wormhole (WH) spacetimes. To this end, we have here performed an in-depth investigation of the general properties and features of braneworld WHs generated by the GEA.

In the original formulation of the GEA~\cite{Nakas:2023yhj}, the seed brane metric is given in terms of the curvature radial coordinate $r$ (MT frame) that is identified with the circumferential radius of the 4D geometry. Our analysis reveals the important role of $r$ in the uplift of any static and spherically symmetric brane spacetime written in terms of an arbitrary radial coordinate $u$ (nMT frame). For nMT brane metrics, the functional form of an~\emph{invertible} circumferential radius metric function $r(u)$, corresponding to the physical radius of the spacetime, is indirectly involved in the definition of the GEA metric. We show that this is a consequence of the proper way to define wormhole structures in models of extra dimensions, that is in terms of a lower cutoff in the domain of the~\emph{bulk} radial coordinate $\rho$. More generally however, $r(u)$ may be non-invertible and a number of known WH and BH spacetimes fall within this class of spacetimes. In such cases, the original formulation of the GEA can no longer be applied, and in order to address this limitation, we have introduced an extended formulation for the GEA (eGEA) that is valid for $r(u)$ with an arbitrary functional form. When $r(u)$ is invertible, the eGEA reduces to the original GEA.

Based on minimalistic requirements for the GEA line element to describe a wormhole structure in the bulk, we have derived the corresponding general constraints on its metric functions for any type of brane radial coordinate i.e. for MT and nMT WHs. We have also discussed in detail how to determine the coordinate location and physical radius in the most general way for the 5D GEA WHs. Furthermore, we have derived a general formula for the flare-out condition of MT braneworld WHs which we have found to be different from the one in the case of flat-extra dimensions. As such, we have demonstrated the non-trivial dependence of the flare-out conditions, and consequently of the wormhole structure, on the nature of the extra dimensions (flat or warped). Also, by utilizing the symmetries of the GEA metric we have demonstrated how to obtain embedding diagram representations of these 5D braneworld WHs.
Moreover, under the assumption that the theories giving rise to the braneworld wormhole solutions respect Eq.~\eqref{eq: act}, we derived the general form of the stress-energy tensor applicable to both thin and thick braneworld scenarios.
In general, this tensor includes off-diagonal components, as shown in Eq.~\eqref{eq:T-gen}.
Interestingly, our analysis revealed that the off-diagonal components of the stress-energy tensor, associated with shear stress, vanish identically only for the RS-II braneworld model. 
This distinctive feature underscores the RS-II model as a special model among braneworld scenarios.

To illustrate our method, we have considered four well-known 4D spacetimes as seed brane metrics, namely, the Casadio-Fabbri-Mazzacurati braneworld WH~\cite{Casadio:2001jg},  the Bronnikov-Kim zero-Schwarzschild mass braneworld WH~\cite{Bronnikov:2002rn}, the Simpson Visser (SV) spacetime~\cite{Simpson:2018tsi}, and the Ellis-Bronnikov (EB) wormhole~\cite{Ellis:1973yv,Bronnikov:1973fh}. The first two examples are of particular importance since they correspond to solutions of the effective 4D braneworld field equations. Consequently, their herein presented 5D bulk uplift provides for the first time in the literature an answer to the question regarding their higher-dimensional structure in the bulk. The SV metric provides a simple, yet illustrative, example of a wormhole with an invertible $r(u)$ that can be uplifted either in terms of the original GEA or the eGEA, and as such provides an ideal set-up for a comparison between the two uplift methods. Finally, the EB spacetime corresponds to a non-isometric (or also oftentimes called throat asymmetric) wormhole with a non-invertible $r(u)$, and its uplift in terms of the eGEA demonstrates the impact that the asymmetry of the seed brane metrics has on the structure of the uplifted WH spacetimes.

Our analysis of the Ricci scalar, energy conditions, flare-out conditions and the embedding diagram representations for the aforementioned examples in both thin and thick braneworld models, leads to the following general conclusions: When the GEA (in its original or extended formulation) is applied to wormhole brane seed metrics, the resultant 5D spacetimes preserve their wormhole structure. In the low bulk-curvature regime, the braneworld wormhole structure exhibits strong similarities with the flat-extra dimensional case, however, when the warping of the extra dimension is strong, the 5D structure exhibits strong deformation along the extra dimension. In all cases however, the WH is localized near the brane and the spacetimes asymptote to AdS$_5$ exponentially fast at a finite distance from the brane. The wormhole hyperthroat corresponds to a $3$-dimensional spacelike hypersurface defining the boundary of spacetime. The amount of deviation of this hypersurface from sphericity is determined by the bulk AdS$_5$ curvature.

Our work, points towards exciting future research directions in wormhole spacetimes in braneworld models. In particular, the violation of the energy conditions for WHs in the RS-II model seems to be a generic result independent of the choice of the seed brane metric. This raises the intriguing question of whether it is possible to consider a different warp function that would lead to wormhole spacetimes satisfying the energy conditions everywhere along the hyperthroat. The stability of braneworld WHs is yet another important area for future investigation and the freedom in choosing the warp function in the GEA provides some flexibility in this quest. Such an endeavor however requires knowledge of the complete bulk field theory that supports these geometries and this can be facilitated by means of conformal transformations as it has been demonstrated in~\cite{Nakas:2023yhj}.

Furthermore, the herein demonstrated significant impact of the warping of the extra dimension on the wormhole structure provides further motivation to extend the GEA to two-brane setups and identify the conditions for wormhole structures in that case. In a recent intriguing proposal~\cite{Dai:2020rnc}, wormhole-like structures in two-brane models have been considered as a means to address the open question regarding a formation mechanism for WHs. Last but not least, motivation is provided for the generalization of the method to the case of multiple warped extra dimensions. The importance of such a generalization would extend beyond wormhole spacetimes with potential implications for the uplift of 4D metrics into models of String Theory.

\acknowledgments
TDP acknowledges the support of the Research Centre for Theoretical Physics and Astrophysics at the Institute of Physics, Silesian University in Opava. TN is supported by IBS under the project code IBS-R018-D3.

\bibliography{References}{}
\bibliographystyle{utphys}
\end{document}